\documentclass[fleqn,usenatbib]{mnras}

\usepackage{newtxtext,newtxmath}
\usepackage[T1]{fontenc}
\usepackage{float}
\usepackage{placeins}
\usepackage{subcaption}

\DeclareRobustCommand{\VAN}[3]{#2}
\let\VANthebibliography\thebibliography
\def\thebibliography{\DeclareRobustCommand{\VAN}[3]{##3}\VANthebibliography}

\usepackage{graphicx}	% Including figure files
\usepackage{amsmath}	% Advanced maths commands
\usepackage{lscape}
\usepackage{multirow}

\usepackage{amssymb,amsmath}
\usepackage{url}
\usepackage{titlesec}
\usepackage{geometry}
\usepackage{soul}
\usepackage{orcidlink} 
\usepackage{xcolor}
\title[Variable magnetic activity in K2 stars]{Investigating magnetic activity cycles in solar-like oscillators using asteroseismic data from the K2 mission}

\author[G. Berloff et al.]{Gleb Berloff$^{1}$\orcidlink{0009-0009-8457-3589},
Anne-Marie Broomhall$^{2}$\thanks{E-mail: a-m.broomhall@warwick.ac.uk}\orcidlink{0000-0002-5209-9378},
George T. Hookway$^{3}$\orcidlink{0009-0002-8134-4026},
Mikkel N. Lund$^{4}$\orcidlink{0000-0001-9214-5642},
\newauthor
Laura Jade Millson$^{2}$\orcidlink{0009-0003-4254-2676},
and Dmitrii Kolotkov$^{2,5}$\orcidlink{0000-0002-0687-6172}
\\
$^{1}$Department of Physics, University of Warwick, Coventry, CV4 7AL, UK\\
$^{2}$Centre for Fusion, Space and Astrophysics, University of Warwick, Coventry, CV4 7AL, UK\\
$^{3}$School of Physics and Astronomy, University of Birmingham, Edgbaston, Birmingham, B15 2TT, UK\\
$^{4}$Stellar Astrophysics Centre, Department of Physics and Astronomy, Aarhus University, Ny Munkegade 120, 8000, Aarhus C, Denmark\\
$^{5}$$^{2}$Engineering Research Institute \lq\lq Ventspils International Radio Astronomy Centre (VIRAC)\rq\rq, Ventspils University of Applied Sciences, Ventspils, LV-3601, Latvia
}

\date{Accepted XXX. Received YYY; in original form ZZZ}

\pubyear{\the\year{}}

\begin{document}
\label{firstpage}
\pagerange{\pageref{firstpage}--\pageref{lastpage}}
\maketitle

% Abstract of the paper
\begin{abstract}
We present the results of an investigation into the possible presence of magnetic activity cycles in stars observed in two observational campaigns by the K2 mission. This study was based on the KEYSTONE asteroseismic sample of solar-like oscillators, which contained 20 stars for which we were able to determine whether the asteroseismic p-mode frequencies varied in time. These frequency shifts ($\delta\nu$) were determined using a cross-correlation method and using the individual mode frequencies, obtained by fitting power spectra. Three stars were found to exhibit $\delta\nu$ larger than their associated errors ($\sigma_{\delta\nu}$) using both methods, while two more stars exhibited $\delta\nu>\sigma_{\delta\nu}$ when the cross correlation was used and a further two stars exhibited $\delta\nu>\sigma_{\delta\nu}$ when the fitted frequencies were used. When considering the whole sample of 20 stars, the amplitude of $\delta\nu$ showed no dependence on the large frequency separation and metallicity. However, $\delta\nu$ was observed to increase with rotation rate and effective temperature. Our sample contained a number of evolved subgiant stars, allowing us to expand the parameter space usually considered when comparing $\delta\nu$ with stellar parameters. While $\delta\nu$ was small for all of the evolved stars, one was found to have $\delta\nu>\sigma_{\delta\nu}$, raising the possibility that these evolved stars may still exhibit variable magnetic activity.
\end{abstract}

% Select between one and six entries from the list of approved keywords.
% Don't make up new ones.

%%%%%%%%%%%%%%%%%%%%%%%%%%%%%%%%%%%%%%%%%%%%%%%%%%

%%%%%%%%%%%%%%%%% BODY OF PAPER %%%%%%%%%%%%%%%%%%

\section{Introduction}
\label{section: Introduction}

Sunspots are pronounced dark features on the surface of the Sun associated with colder-than-average temperatures in the solar photosphere. 
In 1843, following an anomalous 75-year period in the 17th-18th century in which almost no sunspots were observed, an era known as the Maunder minimum \citep[see e.g.][and references therein]{2015LRSP...12....4H}, Heinrich Schwabe discovered that the amount of sunspots visible on the surface of the Sun typically varies with an approximate 11-year periodicity \citep{1844AN.....21..233S}. The beginning of each sunspot cycle is indicated by sunspots appearing at mid-latitudes, after which they are observed at latitudes increasingly close to the equator, forming what is now known as the butterfly diagram \citep{1904MNRAS..64..747M, 2015LRSP...12....4H}. 
The first association that sunspots are related to magnetic activity was made by \citet{1908ApJ....28..315H}, who discovered that sunspots typically arose in pairs of opposite magnetic polarity, and that the apparent solar sunspot cycle of 11\,years is related to a magnetic activity cycle, with a period of 22\,years \citep{Balogh2009}. The magnetic cycle of the Sun is now modelled via a magnetic dynamo typically thought to arise due to turbulent shearing flows in the convective zone of the Sun, although there are still many outstanding questions about exactly how and why the magnetic field of the Sun arises \citep{charbonneau2014solar}.

 Considering that stars are located significantly further away than the Sun, and as such, direct observations of starspots are impractical, many other methods have been devised to infer the presence of activity cycles in other stars. For example, chromospheric indicators of activity have been observed on other stars, most notably as part of the Mount Wilson survey \citep{1995ApJ...438..269B}. Solar values of these activity proxies are known to correlate well with sunspot observations \citep{2024MNRAS.535.2394F}. The Mount Wilson survey revealed the existence of magnetic activity cycles in other stars that, for a single polarity change (akin to our 11\,yr solar cycle), ranged between 1\,yr and 21\,yrs. By studying and analysing the magnetic activity cycles or their indicators in other stars, it is hoped, among other aspects, to formalise constraints on existing dynamo models of stellar magnetic cycles over a broader parameter range. In this study, we use an alternative proxy of stellar activity based upon observations of stellar oscillations.

Observations of the Sun have long suggested that as the activity of the star increases, the frequencies of the p-mode oscillations also increase \citep[][and references therein]{1985Natur.318..449W, 1990Natur.345..322E, broomhall2014sun}. The change in frequency of p modes as a function of time is known as the frequency shift, $\delta\nu$. Observed frequency shifts are commonly thought to be caused either directly by magnetic fields (with the Lorentz force providing a restoring force that leads to an increase in the mode frequency) or via indirect effects, which occur because the magnetic field changes the internal stratification of the star. It is not yet clear which of these effects dominate \citep[e.g.][]{Dziembowski_2005, 2018ApJ...854...74K, 2024MNRAS.533.3387K}.
 
The first asteroseismic evidence for a magnetic activity cycle on a star other than the Sun was on HD~49933, which was observed by the CoRoT mission \citep{2010Sci...329.1032G}. The same behaviour has now been detected in other stars (see \citealt{2023SSRv..219...54J} for a recent review), most notably stars observed by NASA's Kepler mission \citep{2010Sci...327..977B}. In fact, there are now sufficient observations of frequency shifts on stars that it has been possible to investigate how the amplitude of these frequency shifts, and by inference the amplitudes of stellar cycles, may relate to different stellar parameters. \citet{santos2019signatures} compared p-mode frequency shifts detected on 75 stars observed by Kepler to various stellar parameters, finding that the amplitude of the frequency shift correlated positively with the chromospheric activity indicator $R_{\mathrm{HK}}$ and rotation rate, and was anti-correlated with stellar age. \citeauthor{santos2019signatures} also noticed a positive correlation with effective temperature and postulated that this was because the hotter a star, the more sensitive the p modes are to any magnetic perturbations, in agreement with predictions made by \citet{2007MNRAS.379L..16M} and \citet{2019FrASS...6...52K}. However, it must be remembered that even for stars observed during the full $\approx$4 years of the Kepler mission, the observations may not cover a full stellar cycle.

Kepler stared at one patch of the sky for the entirety of its nominal mission (2009--2013). However, when Kepler was no longer able to maintain stable pointing, it was repurposed as K2 \citep{2014PASP..126..398H}, which operated between 2014 and 2018. Unlike Kepler, K2 changed its field of view regularly and made photometric observations of stars in specified areas along the ecliptic, referred to
as ``campaigns''. The standard duration of a K2 campaign was just 80 days, which is not sufficient to determine frequency shifts alone. However, some of the campaigns overlapped, meaning that certain stars were observed in multiple non-consecutive campaigns by K2. Therefore, this study aims to detect frequency shifts and, by inference, magnetic activity cycles, by comparing asteroseismic observations in two different K2 campaigns. Although with only two data points we cannot determine whether any detected shifts represent activity cycles, this study highlights stars worthy of further observation and contributes to our understanding of which stellar parameters play a role in determining the amplitude of activity variations. 

Since we aim to detect evidence for magnetic activity using variations in p-mode frequencies, our focus is on solar-like oscillators. Previous works, such as \citet{santos2018signatures}, have concentrated only on main-sequence solar-like oscillators. However, subgiants (and red giants) are also solar-like oscillators, and we include several subgiants in our sample. While these older stars are not expected to be as magnetically active as their main sequence counterparts \citep{1972ApJ...171..565S, 2008ApJ...687.1264M}, \citet{2020MNRAS.496.4593K} predicted that evolved stars may be more sensitive to any magnetic field and so we also include sub giants in our sample, expanding the parameter space considered by \citet{santos2018signatures, santos2019signatures}. \citet{2025A&A...700A..39K} used asteroseismology to study a subgiant star, $\beta$~Aql, which appears to have a $4.7\pm0.4\,\rm yr$ activity cycle, but found no significant change in frequency between the years 2022 and 2023. This work also represents an extension to \citet{2023A&A...674A.106G} who performed an asteroseismic study of eight stars from K2 campaigns 6 and 17. Here, we do not restrict our study to just these two campaigns; therefore, we expand upon their sample of stars. 

The remainder of the paper is structured as follows. In Section \ref{sec:method} we describe our sample of stars, the data used and the methods by which the frequency shifts were determined. Section \ref{sec:results} contains the main results including a comparison of the observed frequency shifts with various stellar parameters, put into context by comparison with those found by \citet{santos2019signatures}. Section \ref{sec:conclusions} contains our main conclusions.

\section{Method of research}\label{sec:method}

\subsection{Identification of target stars}

\citet{2024A&A...688A..13L} recently published the K2 asteroseismic KEYSTONE sample of 173 dwarf and subgiant stars that exhibited solar-like oscillations in power spectra constructed from short-cadence K2 data prepared for asteroseismic analyses by the K2P$^2$ pipeline \citep{2014MNRAS.445.2698H, 2015ApJ...806...30L}. We identified the stars in the KEYSTONE sample that were observed in two K2 campaigns and used the data prepared using the K2P$^2$ pipeline to determine whether any change in oscillation frequency was observed between campaigns: Two methods were used to determine the observed frequency shifts and these will be described in Sections \ref{sec:cross} and \ref{sec:fitted}. \citet{2024A&A...688A..13L} identified 30 stars that were observed in multiple campaigns, however, for 9 stars, their Table 4 only provides values of $\Delta\nu$ and $\nu_\text{max}$ for single campaigns and/or when the data from the campaigns were combined. This was because the signal to noise in one of the campaigns was not sufficient on its own to obtain these parameters. We therefore excluded those stars from our study. Additionally, we found that for EPIC 24630350, the p-mode signal to noise was not sufficient to allow the frequency shifts to be determined. In total, we identified 20 stars that were observed in more than one campaign and were suitable for this study. We now provide further details on the stars in our sample.

\subsection{Parameters of stars in our sample}\label{sec:stellar_params}

Values of the large separation ($\Delta\nu$, which is the frequency spacing between neighbouring radial orders, $\textit{n}$, for modes of the same angular degree, $\ell$) and the frequency of maximum p mode power ($\nu_{\text{max}}$, which is generally measured as the centre of a Gaussian fitted to the acoustic modes) for the 20 stars considered in our sample are given in Table~\ref{tab:stellar_parameters}. The values of $\Delta\nu$ and $\nu_{\text{max}}$, which were incorporated into our method for determining the frequency shifts, were obtained from \citet{2024A&A...688A..13L}. In \citeauthor{2024A&A...688A..13L}, three different methods were used to obtain values for these global asteroseismic parameters. Here we quote the coefficient of variation values, obtained using the method described in \citet{2019MNRAS.482..616B}. In Section \ref{sec:results}, we compare the obtained frequency shifts to various stellar parameters. The stellar parameters considered include: $\Delta\nu$; effective temperature ($T_\text{eff}$); surface gravity ($\log\,g$); metallicity ([Fe/H]); and rotation period ($P_\text{rot}$). For our sample, the values of these parameters were also obtained from \citeauthor{2024A&A...688A..13L} and are included in Table~\ref{tab:stellar_parameters}. We note that \citeauthor{2024A&A...688A..13L} gave values for the projected rotational velocity, $v\sin i$, rather than $P_\text{rot}$. We therefore derived $P_\text{rot}$ from $v\sin i$, taking $i=90^\circ$, meaning the values given in Table~\ref{tab:stellar_parameters} represent an upper estimate. When converting from $v\sin i$ to rotation period, we calculated radius values using the asteroseismic scaling relations \citep[e.g.][]{2019LRSP...16....4G}.

\begin{table*}
\caption{Stellar parameters for stars in the KEYSTONE sample observed by multiple K2 campaigns. Values were obtained from \citet{2024A&A...688A..13L}. The uncertainty on all metallicity ([Fe/H]) values was 0.101\,dex, while the error on all surface gravity (log$\,g$) values was 0.1\,dex.}
    \centering
    \begin{tabular}{ccccccc}
    \hline
        EPIC & $\Delta\nu$ & $\nu_{\textrm{\scriptsize{max}}}$ & $T_{\textrm{\scriptsize{eff}}}$ & log$\,g$ & [Fe/H] & 
        $P_{\textrm{\scriptsize{rot}}}$\\
         & $(\upmu\rm Hz)$ & $(\upmu\rm Hz)$ & (K) & (cgs;dex) & (dex) & 
        (d)\\
        \hline
        211403248 & $21.5\pm0.5$ & $335\pm7$ & $5020\pm40$ & $3.4$ & $0.009$ & $62\pm8$ \\
        211409560 & $19.2\pm0.4$ & $271\pm7$ & $4980\pm40$ & $3.4$ & $-0.024$ & $66\pm10$\\
        212485100 & $87.4\pm0.6$ & $1803\pm21$ & $6100\pm40$ & $4.2$ & $-0.038$ & $12.8\pm1.0$\\
        212487676 & $76.5\pm0.6$ & $1441\pm15$ & $6020\pm50$ & $4.1$ & $-0.313$ & $18.8\pm2.1$\\
        212617037 & $51.0\pm0.5$ & $886\pm28$ & $6570\pm50$ & $3.9$ & $0.012$ & $3.56\pm0.13$\\
        212683142 & $45.7\pm0.5$ & $798\pm10$ & $5920\pm40$ & $3.9$ & $-0.051$ & $25.0\pm2.5$\\
        212708252 & $132.3\pm0.8$ & $2890\pm29$ &$5550\pm30$ & $4.4$ & $-0.094$ & $53\pm19$ \\
        212709737 & $78.1\pm0.5$ & $1640\pm37$ & $6560\pm50$ & $4.2$ & $-0.139$ & $6.40\pm0.28$\\
        212772187 & $87.9\pm0.6$ & $1846\pm40$ & $6470\pm50$ & $4.2$ & $-0.046$ & $5.89\pm0.26$ \\
        245961434 & $47.3\pm0.5$ & $808\pm12$ & $5980\pm40$ & $3.9$ & $-0.007$ & $16.2\pm1.2$\\
        245972483 & $85.0\pm0.6$ & $1793\pm16$ & $6130\pm40$ & $4.2$ & $-0.204$ & $10.0\pm0.6$ \\
        246154489 & $14.4\pm0.4$ & $189\pm3$ & $4950\pm30$ & $3.2$ & $-0.366$ & $89\pm14$ \\
        246212144 & $53.4\pm0.4$ & $979\pm17$ & $6370\pm50$ & $4.0$ & $-0.040$ & $11.1\pm0.6$ \\
        212478598 & $35.7\pm0.6$ & $528\pm5$ & $5050\pm30$ & $3.6$ & $-0.356$ & $57\pm11$\\
        212509747 & $66.4\pm0.5$ & $1306\pm19$ & $6420\pm40$ & $4.1$ & $-0.266$ & $13.5\pm0.9$\\
        246134147 & $83.6\pm0.6$ & $1727\pm22$ & $6240\pm50$ & $4.2$ & $-0.366$ & $10.1\pm0.6$ \\
        246033065 & $102.0\pm0.7$ & $2320\pm50$ & $4130\pm30$ & $4.3$ & $-0.096$ & $9.2\pm0.6$ \\
        246438837 & $55.7\pm0.5$ & $1071\pm17$ & $6160\pm40$ & $4.0$ & $0.005$ & $12.4\pm0.7$\\
        212516207 & $68.3\pm0.5$ & $1398\pm18$ & $6150\pm50$ & $4.1$ & $0.132$ & $15.0\pm1.1$\\
        246305274 & $46.7\pm0.5$ & $799\pm13$ & $6210\pm40$ & $3.9$ & $-0.416$ & $14.8\pm1.0$\\
        \hline
    \end{tabular}
    \label{tab:stellar_parameters}
\end{table*}

\subsection {Determination of oscillation frequency shifts using cross correlations}\label{sec:cross}

Frequency shifts ($\delta\nu$) were determined using the cross-correlation technique described in \citet{2017A&A...598A..77K}. The process is summarised here for clarity. For each star and each campaign, a power spectrum was obtained from the K2P$^2$ cleaned data using a Lomb-Scargle periodogram. It was important when determining the cross correlation that the power spectra were evaluated at the same frequency locations. Therefore, the frequency array used to evaluate the power spectrum for the first campaign was also used to determine the power spectrum for the second campaign. 

\begin{figure*}
    \centering
    \includegraphics[width=0.45\textwidth]{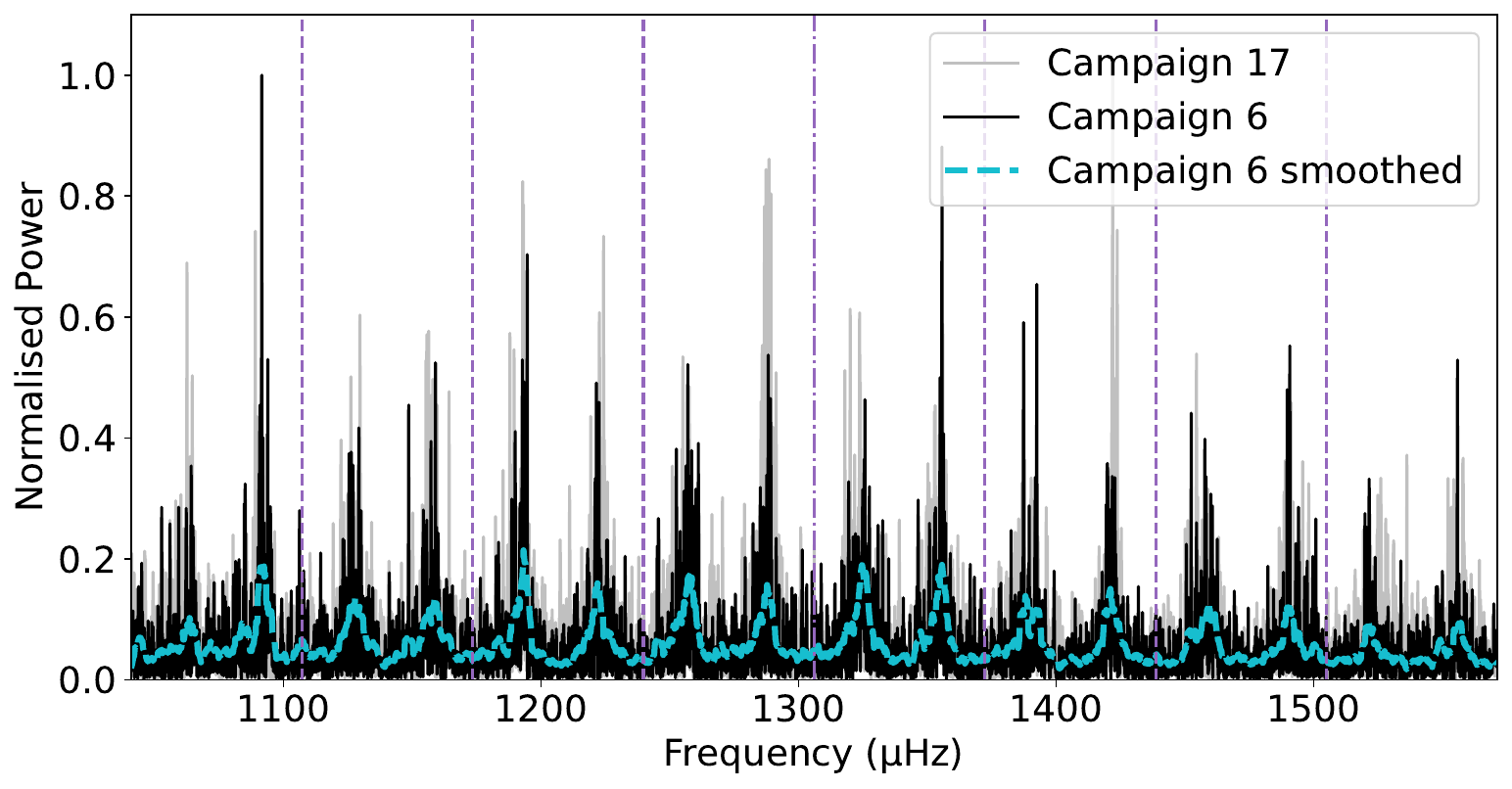}
    \includegraphics[width=0.45\textwidth]{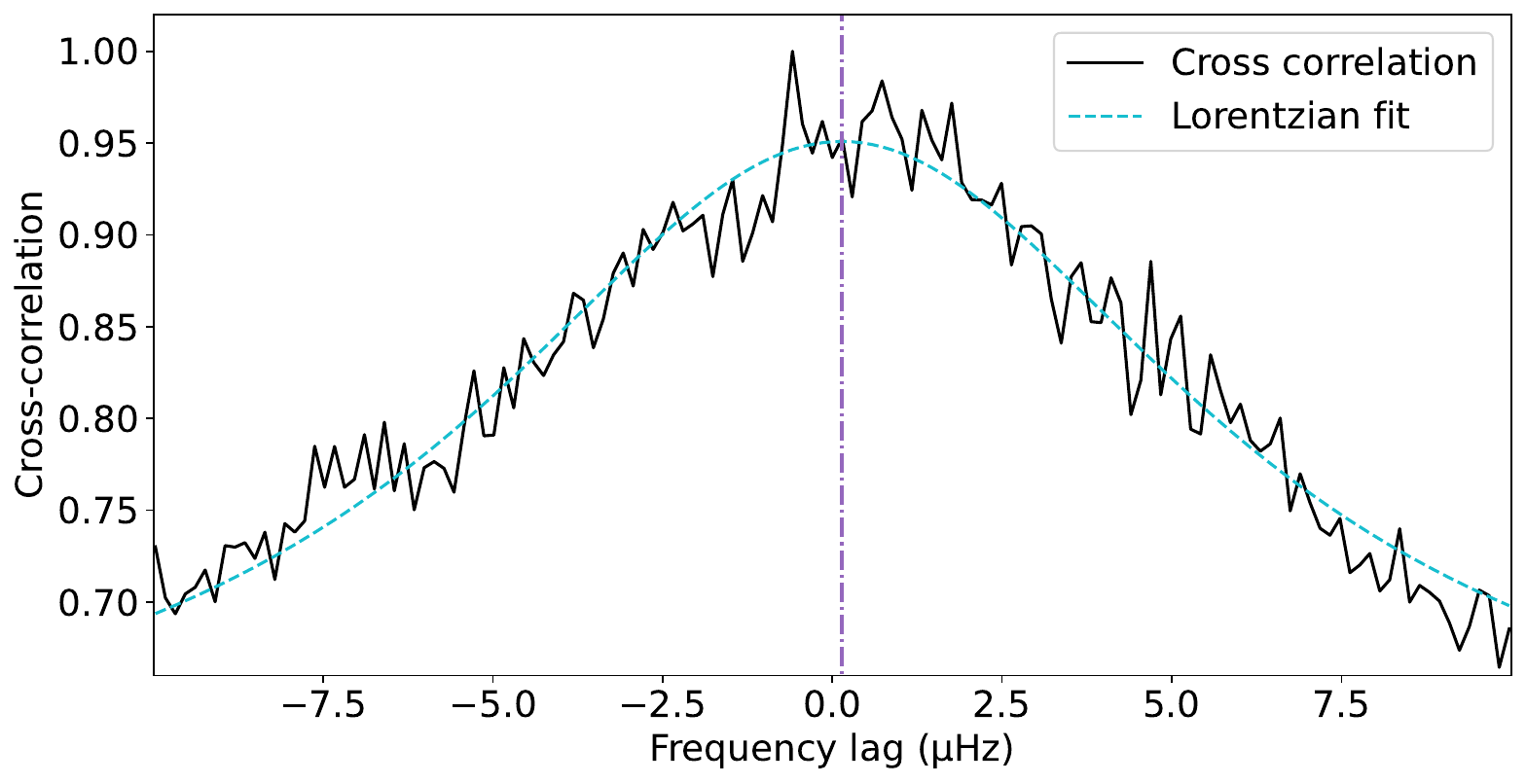}
    \caption{Left: Power spectra observed for EPIC 212509747 in campaigns 6 (black) and 17 (grey), with a smoothed version of campaign 6 (cyan dashed). Here the smoothing was performed using a boxcar of width 22 bins. The spectrum is limited to the frequency range $\nu_{\textrm{max}}\pm4\Delta\nu$. The vertical dot-dashed line indicates $\nu_{\textrm{max}}$, while the vertical dashed lines are separated from $\nu_{\textrm{max}}$ by multiples of $\Delta\nu$. Right: An example of a cross-correlation (black) obtained using a realisation of the power spectra in the left panel, the Lorentzian fitted to the cross correlation (cyan, dashed) and the location of the maximum of the Lorentzian (purple, dot-dashed).}
    \label{fig:example_cross_correlation}
\end{figure*}

Using values of $\nu_{\textrm{\scriptsize{max}}}$ and $\Delta\nu$ from Table \ref{tab:stellar_parameters}, the cross-correlation was determined over a range of $\nu_{\textrm{\scriptsize{max}}}\pm4\Delta\nu$. This range will contain approximately eight overtones for each angular degree that have the highest signal to noise, which is sufficient to determine the frequency shift. A wider range runs the risk of incorporating low signal-to-noise p modes. The left-hand panel of Figure~\ref{fig:example_cross_correlation} shows examples of the spectra over this frequency range that were observed for EPIC 212509747. The cross correlation between these spectra was determined, and the location of the central peak in the cross correlation then corresponds to the change in frequency between the oscillations in the two campaigns. However, since the data in the cross correlation are correlated with each other, rather than fitting the cross correlation itself, 1000 realisations of the power spectra from each campaign were simulated. The realisations were simulated by first smoothing the power spectra obtained from the raw data, using a boxcar smoothing. The smoothed power spectra can be thought of as representing the underlying signal in the data and should, therefore, capture the relatively sharp structure of the p-mode peaks while smoothing out the noise. The width and separation of the mode peaks varied across the sample due to the range of stellar parameters (e.g. $T_\text{eff}$, $\log\,g$) considered. Therefore, the smoothing window, $n_{\textrm{\scriptsize{sm}}}$, was taken to be the maximum of the integer number of frequency bins closest to $0.05\Delta\nu$ or five frequency bins. Following a visual inspection, this was found to represent a good balance between smoothing the noise, while maintaining the structure of the p-mode peaks. A lower limit of five frequency bins was put in place as anything below this would not sufficiently smooth the noise to reveal the signal from the p-mode peaks. For EPIC 212509747 this corresponded to a smoothing width of $n_{\textrm{\scriptsize{sm}}}=22\,\rm bins$ or $3.2\,\rm\upmu Hz$. Although we note that conversion between bins and frequency will vary slightly from star to star, depending on the frequency resolution of the computed Lomb-Scargle periodogram. The square root of the smoothed power spectra was then determined to produce amplitude spectra. We then generated independent `real' and `imaginary' realisations by multiplying the amplitude spectra by arrays of Gaussian distributed random noise, with a mean of zero and standard deviation of unity \citep[see e.g.][]{2017A&A...598A..77K}. The real and imaginary realisation components were then recombined to form a single power spectrum realisation. Realisations were created for each campaign, and the cross correlation of the realisations was determined (see Figure \ref{fig:example_cross_correlation}). The central peak was fitted with a Lorentzian curve over a range of $0\pm10\,\rm\upmu Hz$. A value of $\pm10\,\rm\upmu Hz$ is less than $\Delta\nu$ for all stars in our sample, and thus avoids complications due to multiple peaks in the fitting range but is also far larger than the expected frequency shifts. Visual inspection for all stars confirmed $0\pm10\,\rm\upmu Hz$ was sufficient to allow the main peak to be fitted successfully. This process was repeated 1000 times to produce 1000 realisations of the noise. The observed $\delta\nu$ was then defined as the mean of the fitted peak locations and the uncertainty on this value, $\sigma_{\delta\nu}$, given by the standard deviation.

\subsection{Determination of oscillation frequency shifts using fitted mode frequencies}\label{sec:fitted}
The identification and characterisation of the individual mode frequencies was performed with ``peakbagging'' techniques. Peakbagging is the process of applying models to the power spectrum of a light curve to identify the oscillation peaks, and assigning thema value of the radial order, $n$, angular degree, $\ell$, and azimuthal order, $m$~\citep{Appourchaux_2003}. To perform the peakbagging of the power spectra, we used \texttt{PBjam}~\citep{Nielsen_2021, Nielsen_2025}. The method of using \texttt{PBjam} for fitting the modes for the stars in this sample followed the same method as used in~\citet{Hookway_2025}, and worked as follows.

\texttt{PBjam} performs peakbagging in two main stages, mode identification, followed by a detailed peakbagging. The mode identification began with the application of a background model to the power spectrum, with it consisting of three Harvey profiles~\citep{Harvey_1985} and a shot noise term. These terms model the instrumental variability, rotation, and magnetic activity, as well as convection on the stellar surface. The $\ell=0, 1$ and $2$ modes are characterised by the combination of Lorentzian profiles~\citep{Anderson_1990}, with the frequencies of the individual modes being parameterised with an asymptotic relation. There is increased complexity due to the range in evolutions for the stars in the sample. As stars evolve off the main sequence, the $\ell>0$ modes can become mixed due to the coupling between the p- and g-mode cavities. \texttt{PBjam} has three models for the $\ell=1$ modes to account for this, these being: no coupling; few g modes to many p modes; and many g modes to few p modes. For the no coupling, the asymptotic relation used for the $\ell=2,0$ modes is also used for the $\ell=1$ modes. The other two models involve calculating the strength of the coupling between the p and g modes. \texttt{PBjam} does not model for mixed $\ell=2$ modes due to the mixing of modes reducing with increasing $\ell$ and it becomes computationally expensive for the $\ell=2$ modes. The posterior estimates for the asteroseismic parameters used in the models are calculated using a Bayesian framework. The prior distributions were sampled from a large dataset of thousands of previously analysed stars and stellar models to inform the mode identification posteriors. The posterior probability distributions of the model parameters were sampled using the \texttt{dynesty} nested sampling algorithm~\citep{Speagle_2020}.

The results from the mode identification were passed to the detailed peakbagging stage of \texttt{PBjam}. Here, the models are largely the same as those from the mode identification stage, though now the constraints on the model parameters are relaxed. This allows for effects that would not be captured in the asymptotic relations, such as acoustic glitches and rotational splitting. The priors used on the mode frequencies were $\beta$ distributions, where $\alpha=\beta=5$ and the width of the distribution was set to half the small frequency separation, $\delta\nu_{0,2}$. Normal priors were used for the logarithmic heights and logarithmic widths, with prior widths of 50\%. The posterior sampling for this stage was performed with the \texttt{emcee} Python library~\citep{Foreman_Mackey_2013}, using a Markov chain Monte Carlo algorithm. To validate the results, we computed the ratio between the median absolute deviation of the posterior distribution and the standard deviation of the prior distribution of the frequency results. If this ratio was below $2/3$, it was considered that new information on the frequency of the mode had been gained from the light curve, and the mode was retained. Those above this cut-off were removed from the analysis.

The power spectra used for the peakbagging were provided by the K2P$^2$ pipeline~\citep{2015ApJ...806...30L, 2024A&A...688A..13L}, where the pipeline reduces the effects, such as from systematic variability or potential planetary signals. The pipeline provided power spectra for the individual observational campaigns, as well as a spectrum combining the campaigns, for each star. The combined spectra were used for the mode identification. The identified modes were passed to separate detailed peakbagging analyses, where the power spectra for the individual campaigns were used.

The frequency shifts for individual modes were obtained by subtracting the frequency observed in the second campaign from the frequency observed in the first campaign. We note that this gives the frequency shifts the same sign as those determined by the cross-correlation method. For each star, a weighted average of the obtained individual mode frequency shifts was determined and taken as $\delta\nu$ for this method, with the associated weighted uncertainty giving the error $\sigma_{\delta\nu}$.

\subsection{$S_{\mathrm{ph}}$ as a measure of magnetic activity} \label{sec:sph}

The above methods are based on processing the timeseries of a star to obtain the power spectrum of the data. However, another method of detecting magnetic activity in stars relies on the timeseries itself. As illustrated in \citet{2014A&A...562A.124M}, $S_{\mathrm{ph}}$ is an indicator of magnetic activity, which can be compared between stars, and is defined as the standard deviation of the timeseries. Here, $S_{\mathrm{ph}}$ was evaluated in the manner described in \citet{2014A&A...562A.124M}, applying the correction recommended by \citet{2010ApJ...713L.120J} using the data supplied through the K2P$^2$ pipeline. Since these data have been cleaned for asteroseismic studies, any large deviations in the data have been removed, such as those caused by flares but the statistical noise caused by the presence of small scale active regions (such as plages, faculae, and predominantly starspots, akin to those seen on the Sun) should remain. Using helioseismic data, \citet{2017A&A...608A..87S} demonstrated the $S_\text{ph}$ is well correlated with solar activity proxies, such as sunspot number and Ca HK emission, while other studies \citep[e.g.][]{2016A&A...589A.118S, 2018ApJ...852...46K} have shown similar agreement between $S_{\text{ph}}$ and Ca HK emission for other stars. In \citet{2014A&A...562A.124M}, it was recommended to determine $S_{\mathrm{ph}}$  over a period corresponding to five times the rotation period of the star. This was done to negate variability in $S_\text{ph}$ when determined on shorter timeseries caused by active regions coming in and out of view. \citet{2014A&A...562A.124M} state that as long as $S_\text{ph}$ is determined over a length of time greater than three times $P_\text{rot}$ the value converges towards a constant value. In this study, we aimed to compare the activity levels in two different K2 campaigns and so, for each star, a single value of $S_{\mathrm{ph}}$ was determined for each campaign. Given that the typical length of a K2 campaign is around 80\,d, for most stars, this is more than $3P_\text{rot}$ (see Table \ref{tab:stellar_parameters}). For five of the stars in our sample, this is less than $3P_\text{rot}$ but, since longer timeseries are not available, all we can do is bear this in mind when interpreting results. We then defined a quantity called $\Delta S_{\text{ph}}$, which was the difference between the $S_{\text{ph}}$ values for the two campaigns and thus represents the change in activity levels. Despite the ease of this method, there are limitations to it. For the Sun, the most magnetically active regions are observed near the equator. Assuming a similar distribution of activity, if the star is inclined at an angle to our line of sight, the scatter in the timeseries will be underestimated and, hence, the result will be an underestimate of the $\Delta S_{\text{ph}}$ value.

\subsection{Comparison with Kepler stars}\label{sec:kepler}

In this work, we also aim to examine how our frequency shifts relate to various stellar parameters. As our stellar sample is small, temporal p-mode frequency shifts determined by \citet{santos2018signatures} for an extensive set of Kepler stars were included for comparison. As these shifts were measured for multiple 90-day long segments in time, we must define a single value to characterise the variability of the p-mode frequency shifts. Known as the maximum frequency variation, $\delta\nu_{\mathrm{max}}$, this value provides an estimation for the total amplitude of frequency variation for each Kepler star. 

The method to calculate $\delta\nu_{\mathrm{max}}$ from \citet{santos2019signatures} is outlined as follows. At each time point, a new shift value was randomly drawn according to a Gaussian distribution with a mean equal to the p-mode frequency shift and a standard deviation given by its uncertainty. 
The new values were then smoothed using a 180-day boxcar window, and the range of the resulting series was recorded. This process was repeated 10$^4$ times, with the median of the resulting range values taken as $\delta\nu_{\mathrm{max}}$ and their standard deviation as the associated uncertainty.

The $\delta\nu_{\mathrm{max}}$ was determined for 75  high signal-to-noise (SNR) Kepler stars, as selected by \citet{santos2019signatures}. The authors selected stars by first calculating the SNR of the p modes in each 90-day subseries, defined as the average ratio of the mode heights of the five central orders to the background signal. They then computed the mean SNR across all subseries with a duty cycle above 70\%, and retained only those targets with a final average SNR exceeding one.

It is worth noting that as \citet{santos2018signatures}'s temporal p-mode frequency shifts were calculated relative to a reference segment, we do not know where in the magnetic activity cycle we view or how much of it we observe. Without repeated observations covering the full activity cycle, the true maximum frequency shift for each star cannot be determined.
The $\delta\nu_{\mathrm{max}}$ may, therefore, underestimate the total amplitude of frequency variation.

Frequency shifts determined by this work, along with $\delta\nu_{\mathrm{max}}$ for the 75 high-SNR Kepler stars, were compared to stellar parameters. The results of which are presented in Section \ref{sec:results}.  Except for $\log \Delta S_{\mathrm{ph}}$, whose calculation is described prior, the stellar parameters equivalent to those described in Section \ref{sec:stellar_params}, for the stars from the high-SNR Kepler stars can be found in \citet{santos2018signatures} and references therein. 

\section{Results and Discussion}\label{sec:results}
\subsection{Observed frequency shifts}

\begin{table*}
 \caption{Frequency shifts ($\delta\nu$) obtained by two different methods: ``Cross'' refers to those frequency shifts obtained from the cross correlation and $n_\textrm{\scriptsize{sm}}$ gives the number of bins smoothed over before creating the 1000 realisations; ``Fitted'' refers to the frequency shifts obtained using the fitted mode frequencies. Also included in the table are details of which K2 campaigns were used for each star and $\log \Delta S_{\mathrm{ph}}$, which is a proxy for the change in magnetic activity between the two campaigns. An asterisk indicates that \citet{2023A&A...674A.106G} found no evidence of a frequency shift for these stars, while a dagger indicates that the stars were in the sample of \citet{2023A&A...674A.106G} but that they did not determine whether any frequency shift was observed.}
    \centering
    \begin{tabular}{ccccccc}
    \hline
        EPIC & \multicolumn{2}{c}{Campaign} & \multicolumn{2}{c}{Frequency shift $(\upmu\rm Hz)$}  & $n_{\textrm{\scriptsize{sm}}}$ & $\log \Delta S_{\mathrm{ph}}$  \\
         & 1 & 2 & Cross & Fitted &   \\
    \hline
    \multicolumn{7}{c}{$\delta\nu<\sigma_{\delta\nu}$}\\
    \hline
        211403248 & 16 & 18 & $-0.1\pm0.4$ & $0.10\pm0.12$ & 7 & $2.95$\\
        211409560 & 16 & 18 & $-0.2\pm0.5$ & $0.01\pm0.14$ & 6 & $2.75$\\
        212485100* & 6 & 17 &  $-0.18\pm0.29$ & $-0.11\pm0.15$ & 29 & $3.17$\\
        212487676* & 6 & 17 &  $0.0\pm0.2$ & $0.05\pm0.13$ & 26 & $1.83$\\
        212617037* & 6 & 17 &  $0.1\pm0.5$ & $-$ & 17 & $1.76$\\
        212683142$^\dagger$ & 6 & 17 & $-0.02\pm0.14$ & $0.06\pm0.08$ & 15 & $2.55$\\
        212708252 & 6 & 17 & $-0.3\pm0.5$ & $-0.01\pm0.09$ & 45 & $2.19$\\
        212709737 & 6 & 17 &  $0.0\pm0.8$ & $-0.4\pm0.6$ & 26 & $1.83$\\
        212772187* & 6 & 17 & $0.3\pm0.5$ & $0.1\pm0.3$ & 30 & $2.46$\\
        245961434 & 12 & 19 & $-0.2\pm0.3$ & $0.14\pm0.18$ & 16 & $2.94$\\
        245972483 & 12 & 19 & $-0.2\pm0.3$ & $0.12\pm0.23$ & 28 & $3.10$\\
        246154489 & 12 & 19 & $0.1\pm0.3$ & $-0.11\pm0.19$ & 5 & $2.15$\\
        246212144 & 12 & 19 &  $0.0\pm0.3$ & $0.0\pm0.3$ & 18 & $2.36$\\
        \hline
        \multicolumn{7}{c}{$\delta\nu>\sigma_{\delta\nu}$, both methods}\\
        \hline
        212478598$^\dagger$ & 6 & 17 & $0.13\pm0.12$ & $0.04\pm0.03$ & 11 & $1.06$\\
        212509747* & 6 & 17 & $0.4\pm0.3$ & $0.19\pm0.17$ & 22 & $2.33$\\
        246143147 & 12 & 19 & $1.0\pm0.3$ & $0.4\pm0.3$ & 28 & $3.39$\\
        \hline
        \multicolumn{7}{c}{$\delta\nu>\sigma_{\delta\nu}$, cross only}\\
        \hline
        246033065 & 12 & 19 & $0.9\pm0.7$ & $0.2\pm0.4$ & 35 & $2.78$\\
        246438837 & 12 & 19 & $0.6\pm0.3$ & $0.1\pm0.3$ & 18 & $3.09$\\
        \hline
        \multicolumn{7}{c}{$\delta\nu>\sigma_{\delta\nu}$, fitted only}\\
        \hline
        212516207* & 6 & 17 &  $-0.1\pm0.3$ & $-0.38\pm0.17$ & 23 & $2.32$\\
        246305274 & 12 & 19 & $0.1\pm0.3$ & $-0.25\pm0.22$ & 16 & $2.48$\\
    \hline
    \end{tabular}
    \label{tab:shifts}
\end{table*}

\begin{figure}
    \centering
    \includegraphics[width=0.45\textwidth]{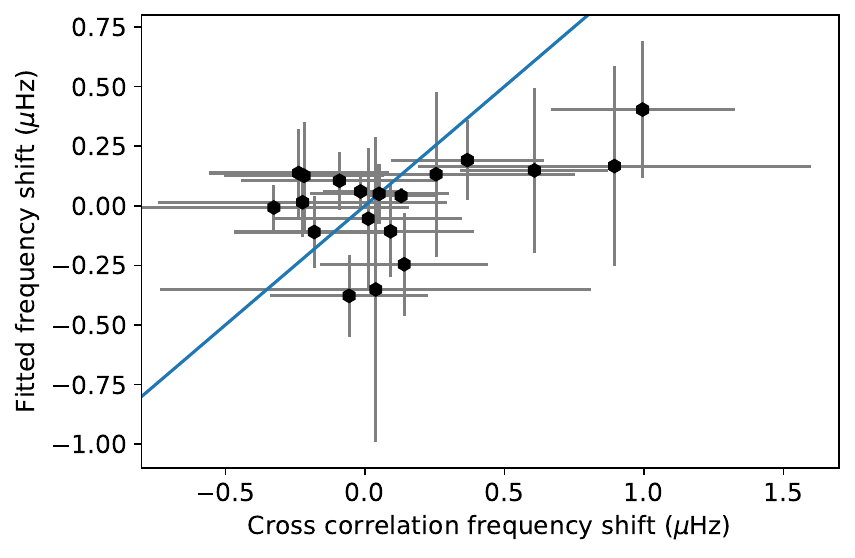}
    \caption{Comparison of the frequency shifts obtained using cross correlations and the fitted frequencies (data points). A 1:1 line is included for guidance.}
    \label{fig:shift_comparison}
\end{figure}

Table \ref{tab:shifts} shows the observed $\delta\nu$ for each star using the cross-correlation method and the fitted mode frequencies. Figure \ref{fig:shift_comparison} shows a comparison between the frequency shifts obtained with the two methods. There is no clear one-to-one trend. However, this was not entirely unexpected given that the frequency shifts were generally small compared to their uncertainties and we note that the difference in the obtained frequency shift by the two methods was less than $1.1\sigma$ for all stars, demonstrating broad agreement. We note that, in agreement with previous studies, the errors associated with the frequency shifts determined using the fitted frequencies were smaller than those obtained from cross correlations. As can be seen in Table \ref{tab:shifts}, no frequency shifts were obtained using the fitted frequencies for EPIC 212617037. This was because no modes passed the fitted validation test in either campaign, indicating a low confidence in the validity of the fitted frequencies, with the power spectrum not being able to provide sufficient new information on the mode compared to the prior.

Three stars exhibit $\delta\nu>\sigma_{\delta\nu}$ when using both the cross-correlation method and the fitted frequencies (EPICs 212478598, 212509747 and 246134147). A further two stars (EPIC 246033065 and EPIC 246438837) exhibit $\delta\nu>\sigma_{\delta\nu}$ when using the cross-correlation method but not the fitted frequencies, while two stars (EPIC 212516207 and EPIC 246305274) were found to have $\delta\nu>\sigma_{\delta\nu}$ when using the fitted frequencies but not the cross-correlation method. Therefore, seven stars in total exhibit $\delta\nu>\sigma_{\delta\nu}$ for at least one method. For the remaining 13 stars, $\delta\nu<\sigma_{\delta\nu}$ for both methods. 

\citet{2023A&A...674A.106G} studied eight stars, all of which are in our sample, observed in campaigns 6 and 17. These stars are highlighted using an asterisk or dagger symbol in Table \ref{tab:shifts}. \citeauthor{2023A&A...674A.106G} did not look for evidence of frequency shifts in EPIC 212478598 and EPIC 212683142 (highlighted using the dagger symbol), since they are sub-dwarfs and therefore are not expected to be magnetically active. However, given the results of \citet{2019FrASS...6...52K}, which suggest that even a small change in the magnetic field could produce noticeable frequency shifts in evolved stars, we left these stars in our sample. We note that EPIC 212478598 was one of the stars in our sample that showed a $\delta\nu > \sigma_{\delta\nu}$ in both methods (while EPIC 212683142 shows no evidence for a frequency shift). \citet{2023A&A...674A.106G} found no evidence of frequency shifts in the remaining six stars. These stars are highlighted by asterisks in Table \ref{tab:shifts}, where it can be seen that we found no evidence of a frequency shift for four of the stars but did find $\delta\nu>\sigma_{\delta\nu}$ (the same criteria as considered by \citeauthor{2023A&A...674A.106G}) for two stars: EPIC 212509747 (using both methods) and EPIC 212516207 (using the fitted frequencies only). Although \citeauthor{2023A&A...674A.106G} also used fitted frequencies to determine whether there were frequency shifts, they used a different fitting code and time series preparation techniques, which may have contributed to this discrepancy.

The fact that the frequency shifts are relatively small for the majority of stars is somewhat expected, given that the difference in time between two overlapping campaigns is relatively small. The start dates of campaigns 16 and 18 are separated by 156\,days, the start dates of campaigns 6 and 17 are separated by 961\,days, and the start dates of campaigns 12 and 19 are separated by 622\,days. These are all relatively short time frames compared to the Sun's approximately 11\,yr magnetic activity cycle.

Even if a star exhibits an activity cycle, there are various factors that would determine whether we would see evidence for that cycle within the data. For instance, the lengths of stellar cycles are typically of the order of years \citep{1995ApJ...438..269B}. While the length of a cycle is observed to increase with increasing rotation period \citep[e.g.][and references therein]{2017SoPh..292..126M}, there is still a large amount of scatter in this relation. The reason for this the scatter is still not well understood \citep[see][for a recent review]{2023SSRv..219...54J}, making it difficult to predict, a priori, what the length of a cycle for a star will be. It is also worth noting that within the Mount Wilson sample, only around 50 per cent of observed stars showed evidence for activity cycle-like behaviour, with a further ${\sim}25$ per cent showing more chaotic variability in activity and ${\sim}25$ per cent showing little to no variability in activity whatsoever \citep{1995ApJ...438..269B}. Furthermore, the phase of the cycle when the observations take place will impact the magnitude of any frequency shift. Ideally, the observations would take place at the start and end of a rising or falling phase of a cycle, respectively, to maximise the frequency shift. However, we are just as likely to make two observations at around cycle maximum or minimum, when the difference in frequency would be minimal. Any frequency shifts observed here can, therefore, be considered as a minimum for the cycle amplitude. 

\subsection{Dependence of frequency shifts on stellar parameters}\label{sub:stellar_params}

\begin{table*}
 \caption{Spearman's rank correlations ($\rho_\text{sp}$) and corresponding p values between the observed frequency shifts, obtained using the cross-correlation (Cross) method and the fitted mode frequencies (Fitted) and various stellar parameters. Correlation coefficients were obtained using the frequency shifts obtained from the K2 data only (K2) and from the combined K2 and Kepler frequency shifts \citep[from][and Section \ref{sec:kepler}]{santos2019signatures}.}
    \centering
    \begin{tabular}{ccccccccc}
    \hline
        & \multicolumn{4}{c}{K2 only} & \multicolumn{4}{c}{K2 and Kepler}   \\
         & \multicolumn{2}{c}{Cross} & \multicolumn{2}{c}{Fitted} &\multicolumn{2}{c}{Cross} & \multicolumn{2}{c}{Fitted} \\
         & $\rho_\text{sp}$ & p value & $\rho_\text{sp}$ & p value & $\rho_\text{sp}$ & p value & $\rho_\text{sp}$ & p value\\
    \hline
    $\Delta\nu$ & 0.421 & 0.065 & 0.257 & 0.286 & 0.112 & 0.395 & 0.059 & 0.655 \\
    $T_\text{eff}$ & 0.120 & 0.613 & 0.678 & 0.001 & 0.431 & $5\times10^{-4}$ & 0.562 & $3\times10^{-6}$\\
    log$\,g$ & 0.395 & 0.084 & 0.254 & 0.293 & 0.137 & 0.296 & 0.084 & 0.525 \\
    $[$Fe/H$]$ & 0.117 & 0.622 & 0.146 & 0.552 & 0.230 & 0.076 & 0.305 & 0.019 \\
    $P_\text{rot}$ & -0.250 & 0.289 & -0.600 & 0.007 & -0.462 & 0.001 & -0.605 & $8\times10^{-6}$\\
    log$\,\Delta S_\text{ph}$ & 0.347 & 0.133 & 0.356 & 0.135 & -0.168 & 0.198 & -0.315 & 0.015\\
    \hline
    \end{tabular}
    \label{tab:spearman}
\end{table*}

\begin{figure*}
    \centering
    \includegraphics[width=0.45\textwidth]{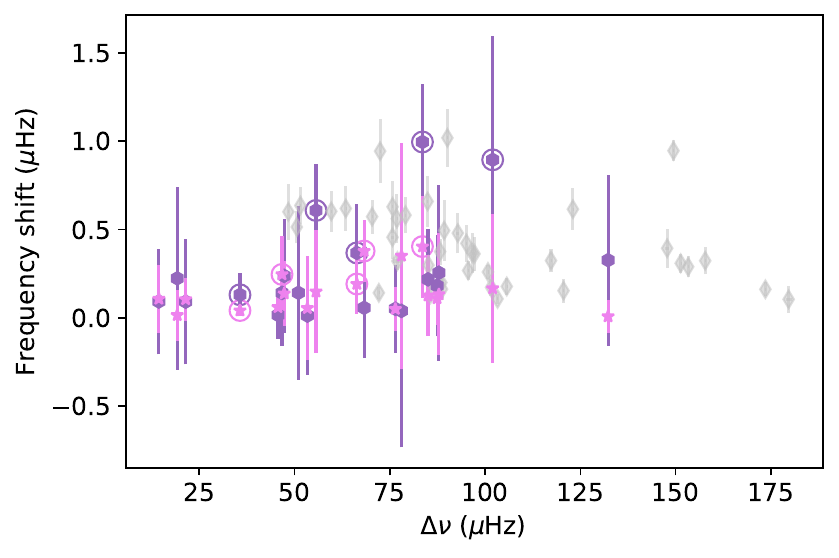}
    \includegraphics[width=0.45\textwidth]{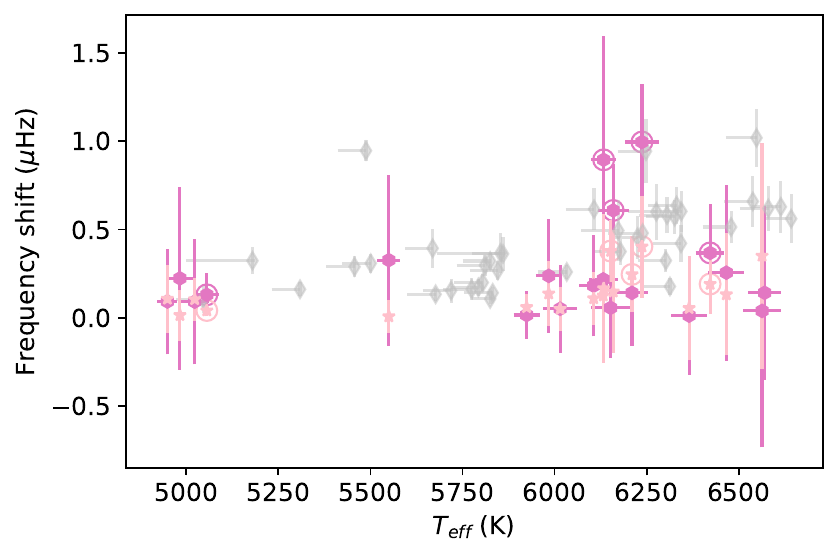}\\
    \includegraphics[width=0.45\textwidth]{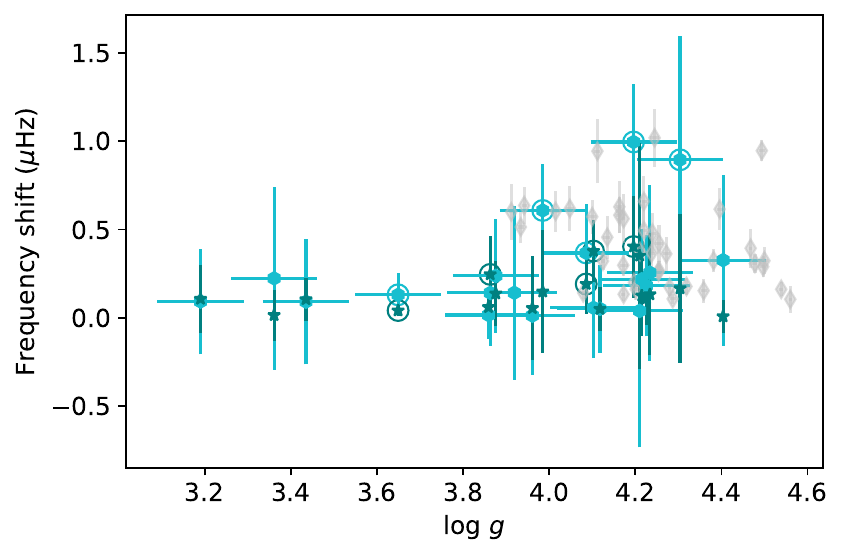}
    \includegraphics[width=0.45\textwidth]{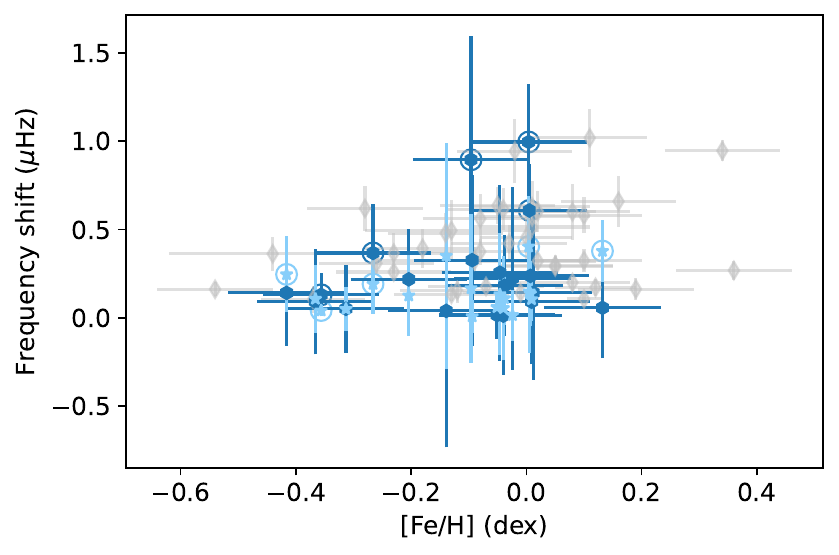}\\
    \includegraphics[width=0.45\textwidth]{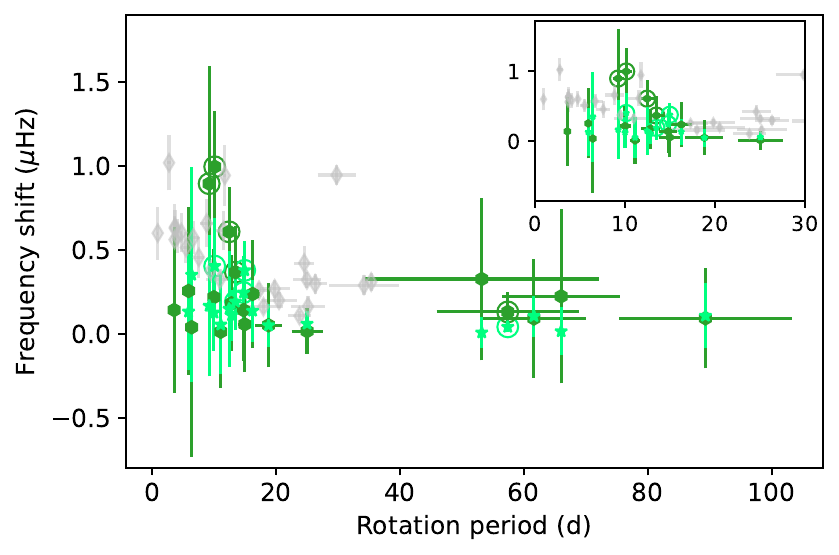}
    \includegraphics[width=0.45\textwidth]{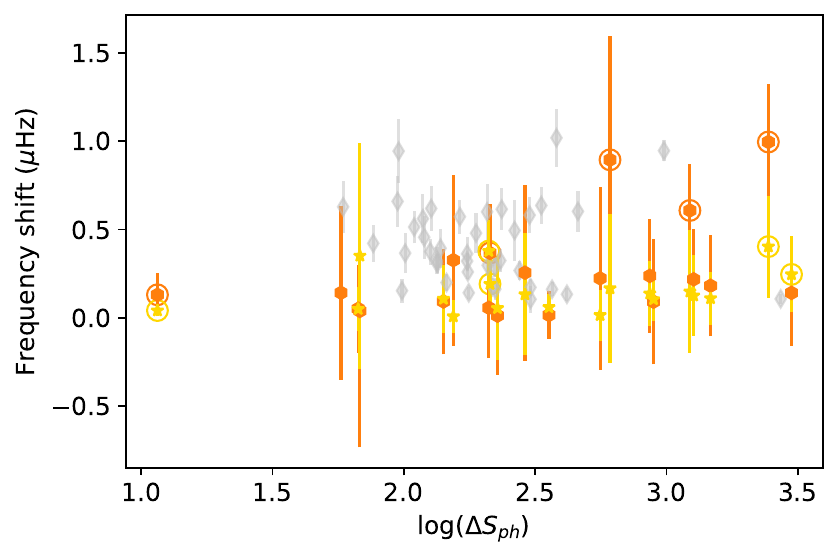}\\
    \caption{Variation in frequency shift with different stellar parameters. In all figures, the hexagon symbols denote results obtained using the cross-correlation method, while the star symbols denote results obtained using the fitted frequencies. Open circles highlight those stars where the observed shift was larger than the uncertainty. All frequency shifts are the absolute values since it is the magnitude of the shift that is important, not whether the frequencies have increased or decreased between campaigns. The grey data were based on the results of \citet{santos2019signatures}. Top left: Large frequency separation, $\Delta\nu$. Top right: Effective temperature, $T_{\mathrm{eff}}$. Middle left: Surface gravity, $\log\,g$. Middle right: Metallicity ([Fe/H]). Bottom left: $\log \Delta S_{\mathrm{ph}}$. Bottom right: Rotation period, $P_{\text{rot}}$, where the inset focuses on the faster rotating stars in our sample. }
    \label{fig:stellar_parameters}
\end{figure*}

Figure \ref{fig:stellar_parameters} demonstrates how the observed frequency shifts vary with different stellar parameters. For comparison, we have also included the results from \citet{santos2019signatures}. Table \ref{tab:spearman} gives the Spearman's rank correlation $\rho_\text{sp}$ and corresponding p values between the frequency shifts and stellar parameters. These were computed using just the K2 frequency shifts and by combining with the Kepler results from \citet{santos2019signatures} (and as described in Section \ref{sec:kepler}).

The top left panel of Figure \ref{fig:stellar_parameters} shows the variation in frequency shift with $\Delta\nu$. It can be seen that our sample extends to lower $\Delta\nu$ values than considered by \citet{santos2019signatures}, as their sample only contained main-sequence stars, whereas ours also contains some subgiant stars. Only one of these evolved stars exhibits frequency shifts larger than their errors. We note that the range of frequency shifts observed is consistent with those seen by \citeauthor{santos2019signatures}, however, the error bars are noticeably larger for the K2 stars. This is likely due to additional noise present in the K2 data compared to the more stable Kepler data. No noticeable trend was observed between $\Delta\nu$ and $\delta\nu$, and the $\rho_\text{sp}$ values were not significant (see Table \ref{tab:spearman}). However, we note that although our sample of stars with $\delta\nu>\sigma_{\delta\nu}$ is only small, the magnitude of the shift for those stars appears to generally increase with $\Delta\nu$. 

The top right panel of Figure \ref{fig:stellar_parameters} shows the variation in $\delta\nu$ with effective temperature, $T_\text{eff}$. \citeauthor{santos2019signatures} observed a positive correlation (Spearman's rank of 0.68) between the frequency shifts and $T_{\mathrm{eff}}$. When considering the frequency shifts obtained using the cross-correlation method, no such correlation was observed, and the observed $\rho_\text{sp}$ was not significant when using only the K2 data. When combined with the Kepler results, the value of $\rho_\text{sp}$ was significant, but lower than that observed by \citeauthor{santos2018signatures}, reflecting the fact that no correlation was observed in the K2 data. However, when the fitted frequencies were used to determine $\delta\nu$ a value of $\rho_{\text{sp}}=0.678$ was observed for just the K2 data, which was significant at a 1\,per cent level. When combined with the Kepler data, the value of $\rho_\text{sp}$ decreased to 0.562 but is more significant, due to the substantial increase in the number of data points. 

The middle left panel of Figure \ref{fig:stellar_parameters} shows how the frequency shift varies with surface gravity (log $\textit{g}$). This result again highlights the fact that the larger shifts in our sample were observed for the main-sequence stars (those with higher log $\textit{g}$ values), while the majority of evolved stars show shifts consistent with zero. Although \citet{santos2019signatures} did not plot the variation in $\delta\nu$ as a function of $\log\,g$, it can be seen from Figure \ref{fig:stellar_parameters} that our sample consists of a wider range of $\log\,g$ values (because of the presence of subgiants).  As can be seen from Figure \ref{fig:stellar_parameters}, the largest shifts in our sample were observed for stars with $\log\,g\approx4.2$, which is broadly in agreement with the results of \citeauthor{santos2019signatures} as plotted here. No significant correlation was observed between the frequency shifts and $\log\,g$.

The middle right panel shows frequency as a function of metallicity ([Fe/H]). While \citeauthor{santos2019signatures} found no significant correlation with metallicity, they did find tentative evidence for two sequences. Again, our results are consistent with this. No significant correlations were observed (see Table \ref{tab:spearman}) but we note the presence of stars with shifts larger than their errors populating the upper sequence identified by \citeauthor{santos2019signatures}. 

The bottom left panel shows the variation in frequency shift with surface rotation period. \citeauthor{santos2019signatures} found a negative Spearman's rank correlation of -0.61 between these two parameters. Our results are consistent with this trend, with the stars that show the largest shifts, i.e. those with $\delta\nu>\sigma_{\delta\nu}$, also being relatively fast rotators. The agreement between these stars and those from \citet{santos2019signatures} can be seen more clearly in the inset figure, which focuses on the faster rotating stars in the sample. We note that some of the fast rotators had frequency shifts consistent with zero. However, as mentioned previously, this may simply be because the time separation of the overlapping campaigns was not large enough. Similarly, those campaigns may coincide with either a maximum or minimum phase in an activity cycle, in which case, the frequencies would remain approximately constant. Table \ref{tab:spearman} demonstrates that while $\rho_\text{sp}$ was found to be negative when considering the K2 frequency shifts obtained using the cross-correlation method, this correlation was not significant, likely due to the frequency shifts on fast rotators being consistent with zero. However, when using the frequency shifts obtained from the fitted frequencies, $\rho_\text{sp}=-0.600$, which was significant at a 1\,per cent level. Our sample spans a wider range of rotation rates than \citeauthor{santos2018signatures}, again reflecting the fact that our sample contains subgiant stars. Our results demonstrate that the expected decrease in frequency shift, and thus by inference magnetic activity, with increasing rotation period extends to higher rotation rates than previously studied. 

The bottom right panel shows the variation in frequency shift with $\log \Delta S_{\mathrm{ph}}$. While no significant correlation was observed (see Table \ref{tab:spearman}), it is worth pointing out that those stars with the largest frequency shifts also have relatively large values of $\log \Delta S_{\mathrm{ph}}$. However, it is interesting to note that the sign of the correlation changed from positive when only the K2 data were used to negative when the K2 data were combined with the Kepler data. An outlier in this sample is the evolved star (EPIC 212478598), which shows a small frequency shift, but one that is greater than its error bars, and also by far the lowest variation in $S_{\mathrm{ph}}$.

\begin{figure}
    \centering
    \includegraphics[width=0.45\textwidth]{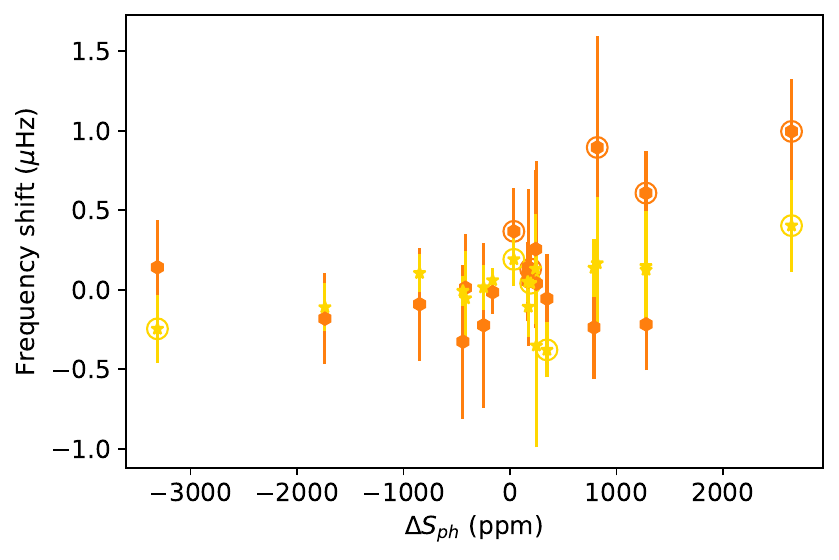}\\
    \caption{Variation in frequency shift with $\Delta S_\text{ph}$. The hexagon symbols denote results obtained using the cross-correlation method, while the star symbols denote results obtained using the fitted frequencies. Open circles highlight those stars where the observed shift was larger than the uncertainties. }
    \label{fig:dsph}
\end{figure}

Figure \ref{fig:dsph} shows how the non-absolute values of frequency shift vary with $S_\text{ph}$. Spearman's rank correlation coefficient between these two variables was found to be $\rho_{\text{sp}}=~0.37$, with a corresponding p value of $0.1$, when the cross-correlation frequency shifts were used, and $\rho_{\text{sp}}=0.5$, with a p value of $0.03$, when the fitted frequency shifts were used. This suggests that there may be a weak correlation, but that the correlation is not significant. This is not entirely surprising given the large number of stars without significant frequency shifts. However, we note that the three stars with large positive frequency shifts found using the cross-correlation method also have large positive changes in $\Delta S_\text{ph}$.

\section{Conclusions}\label{sec:conclusions}
We have searched for evidence for variations in asteroseismic p-mode frequencies in stars observed by the K2 mission in two different time-separated campaigns. We used the KEYSTONE asteroseismic sample \citep{2024A&A...688A..13L} to identify candidates with previously observed p modes and further used the time series optimised for asteroseismology by that study. Our sample consisted of 20 main-sequence and subgiant stars. We used two methods to determine whether any average change in frequency (known as a frequency shift, $\delta\nu$) existed, namely a cross-correlation method and one based on fitting the observed power spectra to determine the individual frequencies of the modes. Out of these 20 stars, seven stars were found to have frequency shifts larger than the associated errors ($\delta\nu>\sigma_{\delta\nu}$) in total: three using both methods; two using the cross-correlation method only, and a different two using the fitted frequencies only. It was not surprising that the majority of stars did not show evidence for a frequency shift, given uncertainties over cycle lengths, amplitudes and phases or even whether a star is likely to exhibit a cycle at all. However, it would be useful to monitor the stars that show some evidence of frequency shifts in the future to determine if they really do have observable magnetic activity cycles. Of particular note is EPIC 212478598, which is a sub-giant star that exhibited $\delta\nu>\sigma_{\delta\nu}$ using both methods. Evolved stars are not expected to be magnetically active, and indeed, this star has a very low $\log(\Delta S_{\text{ph}})$ value but the fact that its frequencies appear to change, hints at the fact that some magnetic variability may remain, even if hidden beneath the surface (where it could impact the p modes, without causing variations in $S_{\text{ph}}$). This result motivates future studies of evolved stars with visible solar-like oscillations, such as those observed with Kepler, which have been neglected by previous studies.

When comparing the observed frequency shifts with stellar parameters, our results are in broad agreement with the results obtained by \citet{santos2019signatures} using Kepler stars. In particular, larger frequency shifts were observed for stars with shorter rotation periods, which likely reflects the fact that these stars are expected to be more magnetically active, and, importantly, have shorter activity cycles \citep{2017SoPh..292..126M}. A significant positive correlation was observed between the frequency shifts obtained using the fitted frequencies and $T_\text{eff}$, which is again in agreement with \citeauthor{santos2019signatures}. There does appear to be some sensitivity in the amplitude of $\delta\nu$ with $\log\,g$, but no significant correlation was observed. The dependence of the frequency shifts on $T_\text{eff}$ (and potentially $\log\,g$) may relate to sensitivity of the modes to a perturbation as predicted by \citet{2007MNRAS.379L..16M, 2019FrASS...6...52K, 2020MNRAS.496.4593K}. The dependence of $\delta\nu$ on [Fe/H] was consistent with the presence of two branches, as highlighted by \citet{santos2019signatures}. Similarly, no clear trend was observed between $|\delta\nu|$ and $\log(\Delta S_{\text{ph}})$, while only a weak correlation was found between $\delta\nu$ and $\Delta S_{\text{ph}}$ (i.e. accounting for the sign of any change). 

The results of this paper highlight the need to increase the number of stars for which we can obtain frequency shifts so as to better understand stellar activity cycles. When observing a star only twice, there is no guarantee that, even if an activity cycle exists, the magnetic activity levels at the times of observation will be different. The observed frequency shift, therefore, will represent a lower limit on the frequency shift between cycle minimum and maximum. We therefore need a larger number of stars to fully understand how stellar parameters impact activity cycles. Furthermore, dedicated long-term monitoring of stars with known frequency variations is needed to determine whether the observed frequency shifts are due to cycles or more chaotic magnetic activity. For these stars in particular, the K2 results could be combined with longer term monitoring to add to our current database of stars with activity cycles and to help our understanding of solar and stellar dynamos. 

\section*{Acknowledgements}
GB wishes to thank Dr. Valentina Ekimova (University of Virginia) and Charlotte Pettett (University of Warwick) for their aid with this research and publication. 
GB and AMB acknowledge the support of the Undergraduate Research Support Scheme (URSS) offered by Warwick University and the Royal Astronomical Society grant for undergraduate research. AMB, DYK and LJM acknowledge support from the Science and Technology Facilities Council (STFC) grant No. ST/X000915/1 and ST/T000252/1. DYK also acknowledge the Latvian Council of Science Project No. lzp2022/1-0017. LJM gratefully acknowledges support from the UK Science and Technology Facilities Council (STFC) grant ST/W507908/1. MNL. acknowledges support from the ESA PRODEX programme.

This study also acknowledges use of Python packages \texttt{NumPy} \citep{harris2020array}, \texttt{Matplotlib} \citep{Hunter2007}, and \texttt{SciPy} \citep{2020SciPy-NMeth}. This work also made use of \texttt{AstroPy}, a community-developed core Python package and an ecosystem of tools and resources for astronomy \citep{astropy:2013, astropy:2018, astropy:2022}.

%%%%%%%%%%%%%%%%%%%%%%%%%%%%%%%%%%%%%%%%%%%%%%%%%%
\section*{Data Availability}
K2 KEYSTONE data available upon request to MNL.

%%%%%%%%%%%%%%%%%%%% REFERENCES %%%%%%%%%%%%%%%%%%

% The best way to enter references is to use BibTeX:

\bibliographystyle{mnras}
\bibliography{K2_shifts_bib} % if your bibtex file is called example.bib

@ARTICLE{2017A&A...598A..77K,
       author = {{Kiefer}, Ren{\'e} and {Schad}, Ariane and {Davies}, Guy and {Roth}, Markus},
        title = "{Stellar magnetic activity and variability of oscillation parameters: An investigation of 24 solar-like stars observed by Kepler}",
      journal = {\aap},
         year = 2017,
        month = feb,
       volume = {598},
          eid = {A77},
        pages = {A77},
          doi = {10.1051/0004-6361/201628469},
archivePrefix = {arXiv},
       eprint = {1611.02029},
 primaryClass = {astro-ph.SR},
       adsurl = {https://ui.adsabs.harvard.edu/abs/2017A&A...598A..77K}
}

@ARTICLE{2014A&A...562A.124M,
       author = {{Mathur}, S. and {Garc{\'\i}a}, R.~A. and {Ballot}, J. and {Ceillier}, T. and {Salabert}, D. and {Metcalfe}, T.~S. and {R{\'e}gulo}, C. and {Jim{\'e}nez}, A. and {Bloemen}, S.},
        title = "{Magnetic activity of F stars observed by Kepler}",
      journal = {\aap},
         year = 2014,
        month = feb,
       volume = {562},
          eid = {A124},
        pages = {A124},
          doi = {10.1051/0004-6361/201322707},
archivePrefix = {arXiv},
       eprint = {1312.6997},
 primaryClass = {astro-ph.SR},
       adsurl = {https://ui.adsabs.harvard.edu/abs/2014A&A...562A.124M}
}

@ARTICLE{2010ApJ...713L.120J,
       author = {{Jenkins}, Jon M. and {Caldwell}, Douglas A. and {Chandrasekaran}, Hema and {Twicken}, Joseph D. and {Bryson}, Stephen T. and {Quintana}, Elisa V. and {Clarke}, Bruce D. and {Li}, Jie and {Allen}, Christopher and {Tenenbaum}, Peter and {Wu}, Hayley and {Klaus}, Todd C. and {Van Cleve}, Jeffrey and {Dotson}, Jessie A. and {Haas}, Michael R. and {Gilliland}, Ronald L. and {Koch}, David G. and {Borucki}, William J.},
        title = "{Initial Characteristics of Kepler Long Cadence Data for Detecting Transiting Planets}",
      journal = {\apjl},
         year = 2010,
        month = apr,
       volume = {713},
       number = {2},
        pages = {L120-L125},
          doi = {10.1088/2041-8205/713/2/L120},
archivePrefix = {arXiv},
       eprint = {1001.0256},
 primaryClass = {astro-ph.EP},
       adsurl = {https://ui.adsabs.harvard.edu/abs/2010ApJ...713L.120J}
}

@ARTICLE{2023A&A...674A.106G,
       author = {{Gonz{\'a}lez-Cuesta}, L. and {Mathur}, S. and {Garc{\'\i}a}, R.~A. and {P{\'e}rez Hern{\'a}ndez}, F. and {Delsanti}, V. and {Breton}, S.~N. and {Hedges}, C. and {Jim{\'e}nez}, A. and {Della Gaspera}, A. and {El-Issami}, M. and {Fox}, V. and {Godoy-Rivera}, D. and {Pitot}, S. and {Proust}, N.},
        title = "{Multi-campaign asteroseismic analysis of eight solar-like pulsating stars observed by the K2 mission}",
      journal = {\aap},
         year = 2023,
        month = jun,
       volume = {674},
          eid = {A106},
        pages = {A106},
          doi = {10.1051/0004-6361/202244577},
archivePrefix = {arXiv},
       eprint = {2304.00087},
 primaryClass = {astro-ph.SR},
       adsurl = {https://ui.adsabs.harvard.edu/abs/2023A&A...674A.106G}
}

@Inbook{Balogh2009,
author="Balogh, A.
and Thompson, M. J.",
editor="Thompson, M. J.
and Balogh, A.
and Culhane, J. L.
and Nordlund, {\AA}.
and Solanki, S. K.
and Zahn, J.-P.",
title="Introduction to Solar Magnetism: The Early Years",
bookTitle="The Origin and Dynamics of Solar Magnetism",
year="2009",
publisher="Springer New York",
address="New York, NY",
pages="1--14",
isbn="978-1-4419-0239-9",
doi="10.1007/978-1-4419-0239-9_1",
url="https://doi.org/10.1007/978-1-4419-0239-9_1"
}

@article{charbonneau2014solar,
  title={Solar dynamo theory},
  author={Charbonneau, Paul},
  journal={Annual Review of Astronomy and Astrophysics},
  volume={52},
  number={1},
  pages={251--290},
  year={2014},
  publisher={Annual Reviews}
}

@article{broomhall2014sun,
  title={The Sun’s interior structure and dynamics, and the solar cycle},
  author={Broomhall, A-M and Chatterjee, P and Howe, R and Norton, AA and Thompson, MJ},
  journal={Space Science Reviews},
  volume={186},
  pages={191--225},
  year={2014},
  publisher={Springer}
}

@article{Dziembowski_2005,
doi = {10.1086/429712},
url = {https://dx.doi.org/10.1086/429712},
year = {2005},
month = {may},
publisher = {},
volume = {625},
number = {1},
pages = {548},
author = {W. A. Dziembowski and P. R. Goode},
title = {Sources of Oscillation Frequency Increase with Rising Solar Activity},
journal = {The Astrophysical Journal}
}

@article{santos2019signatures,
  title={Signatures of magnetic activity: on the relation between stellar properties and p-mode frequency variations},
  author={Santos, ARG and Campante, TL and Chaplin, WJ and Cunha, MS and Van Saders, JL and Karoff, C and Metcalfe, TS and Mathur, S and Garc{\'\i}a, RA and Lund, MN and others},
  journal={The Astrophysical Journal},
  volume={883},
  number={1},
  pages={65},
  year={2019},
  publisher={IOP Publishing}
}

@ARTICLE{2024A&A...688A..13L,
       author = {{Lund}, Mikkel N. and {Basu}, Sarbani and {Bieryla}, Allyson and {Casagrande}, Luca and {Huber}, Daniel and {Hekker}, Saskia and {Viani}, Lucas and {Davies}, Guy R. and {Campante}, Tiago L. and {Chaplin}, William J. and {Serenelli}, Aldo M. and {Joel Ong}, J.~M. and {Ball}, Warrick H. and {Stokholm}, Amalie and {Bellinger}, Earl P. and {Bazot}, Micha{\"e}l and {Stello}, Dennis and {Latham}, David W. and {White}, Timothy R. and {Sayeed}, Maryum and {B{\o}rsen-Koch}, V{\'\i}ctor Aguirre and {Chontos}, Ashley},
        title = "{The K2 Asteroseismic KEYSTONE sample of Dwarf and Subgiant Solar-Like Oscillators. I. Data and Asteroseismic parameters}",
      journal = {\aap},
         year = 2024,
        month = aug,
       volume = {688},
          eid = {A13},
        pages = {A13},
          doi = {10.1051/0004-6361/202450055},
archivePrefix = {arXiv},
       eprint = {2405.15919},
 primaryClass = {astro-ph.SR},
       adsurl = {https://ui.adsabs.harvard.edu/abs/2024A&A...688A..13L}
}

@ARTICLE{2015ApJ...806...30L,
       author = {{Lund}, Mikkel N. and {Handberg}, Rasmus and {Davies}, Guy R. and {Chaplin}, William J. and {Jones}, Caitlin D.},
        title = "{K2P$^{2}${\textemdash} A Photometry Pipeline for the K2 Mission}",
      journal = {\apj},
         year = 2015,
        month = jun,
       volume = {806},
       number = {1},
          eid = {30},
        pages = {30},
          doi = {10.1088/0004-637X/806/1/30},
archivePrefix = {arXiv},
       eprint = {1504.05199},
 primaryClass = {astro-ph.SR},
       adsurl = {https://ui.adsabs.harvard.edu/abs/2015ApJ...806...30L}
}

@ARTICLE{2014MNRAS.445.2698H,
       author = {{Handberg}, R. and {Lund}, M.~N.},
        title = "{Automated preparation of Kepler time series of planet hosts for asteroseismic analysis}",
      journal = {\mnras},
         year = 2014,
        month = dec,
       volume = {445},
       number = {3},
        pages = {2698-2709},
          doi = {10.1093/mnras/stu1823},
archivePrefix = {arXiv},
       eprint = {1409.1366},
 primaryClass = {astro-ph.IM},
       adsurl = {https://ui.adsabs.harvard.edu/abs/2014MNRAS.445.2698H}
}

@ARTICLE{2014PASP..126..398H,
       author = {{Howell}, Steve B. and {Sobeck}, Charlie and {Haas}, Michael and {Still}, Martin and {Barclay}, Thomas and {Mullally}, Fergal and {Troeltzsch}, John and {Aigrain}, Suzanne and {Bryson}, Stephen T. and {Caldwell}, Doug and {Chaplin}, William J. and {Cochran}, William D. and {Huber}, Daniel and {Marcy}, Geoffrey W. and {Miglio}, Andrea and {Najita}, Joan R. and {Smith}, Marcie and {Twicken}, J.~D. and {Fortney}, Jonathan J.},
        title = "{The K2 Mission: Characterization and Early Results}",
      journal = {\pasp},
         year = 2014,
        month = apr,
       volume = {126},
       number = {938},
        pages = {398},
          doi = {10.1086/676406},
archivePrefix = {arXiv},
       eprint = {1402.5163},
 primaryClass = {astro-ph.IM},
       adsurl = {https://ui.adsabs.harvard.edu/abs/2014PASP..126..398H}
}

@ARTICLE{2019MNRAS.482..616B,
       author = {{Bell}, Keaton J. and {Hekker}, Saskia and {Kuszlewicz}, James S.},
        title = "{Coefficients of variation for detecting solar-like oscillations}",
      journal = {\mnras},
         year = 2019,
        month = jan,
       volume = {482},
       number = {1},
        pages = {616-625},
          doi = {10.1093/mnras/sty2731},
archivePrefix = {arXiv},
       eprint = {1809.09135},
 primaryClass = {astro-ph.SR},
       adsurl = {https://ui.adsabs.harvard.edu/abs/2019MNRAS.482..616B}
}

@ARTICLE{2019LRSP...16....4G,
       author = {{Garc{\'\i}a}, Rafael A. and {Ballot}, J{\'e}r{\^o}me},
        title = "{Asteroseismology of solar-type stars}",
      journal = {Living Reviews in Solar Physics},
         year = 2019,
        month = dec,
       volume = {16},
       number = {1},
          eid = {4},
        pages = {4},
          doi = {10.1007/s41116-019-0020-1},
archivePrefix = {arXiv},
       eprint = {1906.12262},
 primaryClass = {astro-ph.SR},
       adsurl = {https://ui.adsabs.harvard.edu/abs/2019LRSP...16....4G}
}

@ARTICLE{2017SoPh..292..126M,
       author = {{Metcalfe}, Travis S. and {van Saders}, Jennifer},
        title = "{Magnetic Evolution and the Disappearance of Sun-Like Activity Cycles}",
      journal = {\solphys},
         year = 2017,
        month = sep,
       volume = {292},
       number = {9},
          eid = {126},
        pages = {126},
          doi = {10.1007/s11207-017-1157-5},
archivePrefix = {arXiv},
       eprint = {1705.09668},
 primaryClass = {astro-ph.SR},
       adsurl = {https://ui.adsabs.harvard.edu/abs/2017SoPh..292..126M}
}

@ARTICLE{1908ApJ....28..315H,
       author = {{Hale}, George E.},
        title = "{On the Probable Existence of a Magnetic Field in Sun-Spots}",
      journal = {\apj},
         year = 1908,
        month = nov,
       volume = {28},
        pages = {315},
          doi = {10.1086/141602},
       adsurl = {https://ui.adsabs.harvard.edu/abs/1908ApJ....28..315H}
}

@ARTICLE{1844AN.....21..233S,
       author = {{Schwabe}, Heinrich},
        title = "{Sonnenbeobachtungen im Jahre 1843. Von Herrn Hofrath Schwabe in Dessau}",
      journal = {Astronomische Nachrichten},
         year = 1844,
        month = feb,
       volume = {21},
       number = {15},
        pages = {233},
          doi = {10.1002/asna.18440211505},
       adsurl = {https://ui.adsabs.harvard.edu/abs/1844AN.....21..233S}
}

@ARTICLE{2015LRSP...12....4H,
       author = {{Hathaway}, David H.},
        title = "{The Solar Cycle}",
      journal = {Living Reviews in Solar Physics},
         year = 2015,
        month = dec,
       volume = {12},
       number = {1},
          eid = {4},
        pages = {4},
          doi = {10.1007/lrsp-2015-4},
archivePrefix = {arXiv},
       eprint = {1502.07020},
 primaryClass = {astro-ph.SR},
       adsurl = {https://ui.adsabs.harvard.edu/abs/2015LRSP...12....4H}
}

@ARTICLE{1904MNRAS..64..747M,
       author = {{Maunder}, E. Walter},
        title = "{Note on the Distribution of Sun-spots in Heliographic Latitude, 1874-1902}",
      journal = {\mnras},
         year = 1904,
        month = jun,
       volume = {64},
        pages = {747-761},
          doi = {10.1093/mnras/64.8.747},
       adsurl = {https://ui.adsabs.harvard.edu/abs/1904MNRAS..64..747M}
}

@ARTICLE{1995ApJ...438..269B,
       author = {{Baliunas}, S.~L. and {Donahue}, R.~A. and {Soon}, W.~H. and {Horne}, J.~H. and {Frazer}, J. and {Woodard-Eklund}, L. and {Bradford}, M. and {Rao}, L.~M. and {Wilson}, O.~C. and {Zhang}, Q. and {Bennett}, W. and {Briggs}, J. and {Carroll}, S.~M. and {Duncan}, D.~K. and {Figueroa}, D. and {Lanning}, H.~H. and {Misch}, T. and {Mueller}, J. and {Noyes}, R.~W. and {Poppe}, D. and {Porter}, A.~C. and {Robinson}, C.~R. and {Russell}, J. and {Shelton}, J.~C. and {Soyumer}, T. and {Vaughan}, A.~H. and {Whitney}, J.~H.},
        title = "{Chromospheric Variations in Main-Sequence Stars. II.}",
      journal = {\apj},
         year = 1995,
        month = jan,
       volume = {438},
        pages = {269},
          doi = {10.1086/175072},
       adsurl = {https://ui.adsabs.harvard.edu/abs/1995ApJ...438..269B}
}

@ARTICLE{2024MNRAS.535.2394F,
       author = {{Ferreira}, R.~R. and {Gon{\c{c}}alves}, B.~F.~O. and {do Nascimento}, J. -D. and {Castro}, M.},
        title = "{A correlation between sunspot observations and solar Ca II H\&K activity proxies}",
      journal = {\mnras},
         year = 2024,
        month = dec,
       volume = {535},
       number = {3},
        pages = {2394-2403},
          doi = {10.1093/mnras/stae2381},
       adsurl = {https://ui.adsabs.harvard.edu/abs/2024MNRAS.535.2394F}
}

@ARTICLE{1985Natur.318..449W,
       author = {{Woodard}, Martin F. and {Noyes}, Robert W.},
        title = "{Change of solar oscillation eigenfrequencies with the solar cycle}",
      journal = {\nat},
         year = 1985,
        month = dec,
       volume = {318},
       number = {6045},
        pages = {449-450},
          doi = {10.1038/318449a0},
       adsurl = {https://ui.adsabs.harvard.edu/abs/1985Natur.318..449W}
}

@ARTICLE{1990Natur.345..322E,
       author = {{Elsworth}, Y. and {Howe}, R. and {Isaak}, G.~R. and {McLeod}, C.~P. and {New}, R.},
        title = "{Variation of low-order acoustic solar oscillations over the solar cycle}",
      journal = {\nat},
         year = 1990,
        month = may,
       volume = {345},
       number = {6273},
        pages = {322-324},
          doi = {10.1038/345322a0},
       adsurl = {https://ui.adsabs.harvard.edu/abs/1990Natur.345..322E}
}

@ARTICLE{2023SSRv..219...54J,
       author = {{Jeffers}, Sandra V. and {Kiefer}, Ren{\'e} and {Metcalfe}, Travis S.},
        title = "{Stellar Activity Cycles}",
      journal = {\ssr},
         year = 2023,
        month = oct,
       volume = {219},
       number = {7},
          eid = {54},
        pages = {54},
          doi = {10.1007/s11214-023-01000-x},
archivePrefix = {arXiv},
       eprint = {2309.14138},
 primaryClass = {astro-ph.SR},
       adsurl = {https://ui.adsabs.harvard.edu/abs/2023SSRv..219...54J}
}

@ARTICLE{2024MNRAS.533.3387K,
       author = {{Kolotkov}, Dmitrii Y. and {Broomhall}, Anne-Marie and {Hasanzadeh}, Amir},
        title = "{Effects of the photospheric cut-off on the p-mode frequency stability}",
      journal = {\mnras},
         year = 2024,
        month = sep,
       volume = {533},
       number = {3},
        pages = {3387-3394},
          doi = {10.1093/mnras/stae2015},
archivePrefix = {arXiv},
       eprint = {2408.11120},
 primaryClass = {astro-ph.SR},
       adsurl = {https://ui.adsabs.harvard.edu/abs/2024MNRAS.533.3387K}
}

@ARTICLE{2018ApJ...854...74K,
       author = {{Kiefer}, Ren{\'e} and {Roth}, Markus},
        title = "{The Effect of Toroidal Magnetic Fields on Solar Oscillation Frequencies}",
      journal = {\apj},
         year = 2018,
        month = feb,
       volume = {854},
       number = {1},
          eid = {74},
        pages = {74},
          doi = {10.3847/1538-4357/aaa3f7},
archivePrefix = {arXiv},
       eprint = {1801.07932},
 primaryClass = {astro-ph.SR},
       adsurl = {https://ui.adsabs.harvard.edu/abs/2018ApJ...854...74K}
}

@ARTICLE{2010Sci...329.1032G,
       author = {{Garc{\'\i}a}, Rafael A. and {Mathur}, Savita and {Salabert}, David and {Ballot}, J{\'e}r{\^o}me and {R{\'e}gulo}, Clara and {Metcalfe}, Travis S. and {Baglin}, Annie},
        title = "{CoRoT Reveals a Magnetic Activity Cycle in a Sun-Like Star}",
      journal = {Science},
         year = 2010,
        month = aug,
       volume = {329},
       number = {5995},
        pages = {1032},
          doi = {10.1126/science.1191064},
archivePrefix = {arXiv},
       eprint = {1008.4399},
 primaryClass = {astro-ph.SR},
       adsurl = {https://ui.adsabs.harvard.edu/abs/2010Sci...329.1032G}
}

@ARTICLE{2007MNRAS.379L..16M,
       author = {{Metcalfe}, T.~S. and {Dziembowski}, W.~A. and {Judge}, P.~G. and {Snow}, M.},
        title = "{Asteroseismic signatures of stellar magnetic activity cycles}",
      journal = {\mnras},
         year = 2007,
        month = jul,
       volume = {379},
       number = {1},
        pages = {L16-L20},
          doi = {10.1111/j.1745-3933.2007.00325.x},
archivePrefix = {arXiv},
       eprint = {0704.1606},
 primaryClass = {astro-ph},
       adsurl = {https://ui.adsabs.harvard.edu/abs/2007MNRAS.379L..16M}
}

@ARTICLE{2019FrASS...6...52K,
       author = {{Kiefer}, Ren{\'e} and {Broomhall}, Anne-Marie and {Ball}, Warrick H.},
        title = "{Seismic Signatures of Stellar Magnetic Activity - What Can We Expect from TESS?}",
      journal = {Frontiers in Astronomy and Space Sciences},
         year = 2019,
        month = jul,
       volume = {6},
          eid = {52},
        pages = {52},
          doi = {10.3389/fspas.2019.00052},
archivePrefix = {arXiv},
       eprint = {1908.01191},
 primaryClass = {astro-ph.SR},
       adsurl = {https://ui.adsabs.harvard.edu/abs/2019FrASS...6...52K}
}

@Article{         harris2020array,
 title         = {Array programming with {NumPy}},
 author        = {Charles R. Harris and K. Jarrod Millman and St{\'{e}}fan J.
                 van der Walt and Ralf Gommers and Pauli Virtanen and David
                 Cournapeau and Eric Wieser and Julian Taylor and Sebastian
                 Berg and Nathaniel J. Smith and Robert Kern and Matti Picus
                 and Stephan Hoyer and Marten H. van Kerkwijk and Matthew
                 Brett and Allan Haldane and Jaime Fern{\'{a}}ndez del
                 R{\'{i}}o and Mark Wiebe and Pearu Peterson and Pierre
                 G{\'{e}}rard-Marchant and Kevin Sheppard and Tyler Reddy and
                 Warren Weckesser and Hameer Abbasi and Christoph Gohlke and
                 Travis E. Oliphant},
 year          = {2020},
 month         = sep,
 journal       = {Nature},
 volume        = {585},
 number        = {7825},
 pages         = {357--362},
 publisher     = {Springer Science and Business Media {LLC}},
adsurl = {https://www.nature.com/articles/s41586-020-2649-2}
}

@Article{Hunter2007,
  Author    = {Hunter, J. D.},
  Title     = {Matplotlib: A 2D graphics environment},
  Journal   = {Computing in Science \& Engineering},
  Volume    = {9},
  Number    = {3},
  Pages     = {90--95},
  publisher = {IEEE COMPUTER SOC},
  year      = {2007},
    adsurl={https://ieeexplore.ieee.org/document/4160265}
}

@ARTICLE{2020SciPy-NMeth,
  author  = {Virtanen, Pauli},
  title   = {{{SciPy} 1.0: Fundamental Algorithms for Scientific
            Computing in Python}},
  journal = {Nature Methods},
  year    = {2020},
  volume  = {17},
  pages   = {261--272},
    adsurl={https://www.nature.com/articles/s41592-019-0686-2}
}

@article{astropy:2013,
    Archiveprefix = {arXiv},
    Author = {{Astropy Collaboration}},
    Eid = {A33},
    Eprint = {1307.6212},
    Journal = {Astronomy and Astrophysics},
    Month = oct,
    Pages = {A33},
    Primaryclass = {astro-ph.IM},
    Title = {{Astropy: A community Python package for astronomy}},
    Volume = 558,
    Year = 2013,
     adsurl={https://www.aanda.org/articles/aa/full_html/2013/10/aa22068-13/aa22068-13.html}
    
    }

@ARTICLE{astropy:2018,
       author = {{Astropy Collaboration}},
        title = "{The Astropy Project: Building an Open-science Project and Status of the v2.0 Core Package}",
      journal = {The Astronomical Journal},
         year = 2018,
        month = sep,
       volume = {156},
       number = {3},
          eid = {123},
        pages = {123},
archivePrefix = {arXiv},
       eprint = {1801.02634},
 primaryClass = {astro-ph.IM},
    adsurl={https://ui.adsabs.harvard.edu/abs/2018AJ....156..123A/abstract}
}

@ARTICLE{astropy:2022,
       author = {{Astropy Collaboration}},
        title = "{The Astropy Project: Sustaining and Growing a Community-oriented Open-source Project and the Latest Major Release (v5.0) of the Core Package}",
      journal = {Astrophysical Journal},
         year = 2022,
        month = aug,
       volume = {935},
       number = {2},
          eid = {167},
        pages = {167},
archivePrefix = {arXiv},
       eprint = {2206.14220},
 primaryClass = {astro-ph.IM},
adsurl={https://ui.adsabs.harvard.edu/abs/2022ApJ...935..167A/abstract}
}

@ARTICLE{2020MNRAS.496.4593K,
       author = {{Kiefer}, Ren{\'e} and {Broomhall}, Anne-Marie},
        title = "{Empirical relations for the sensitivities of solar-like oscillations to magnetic perturbations}",
      journal = {\mnras},
         year = 2020,
        month = aug,
       volume = {496},
       number = {4},
        pages = {4593-4605},
          doi = {10.1093/mnras/staa1807},
archivePrefix = {arXiv},
       eprint = {2006.11058},
 primaryClass = {astro-ph.SR},
       adsurl = {https://ui.adsabs.harvard.edu/abs/2020MNRAS.496.4593K}
}

@article{santos2018signatures,
  title={Signatures of magnetic activity in the seismic data of solar-type stars observed by Kepler},
  author={Santos, ARG and Campante, TL and Chaplin, WJ and Cunha, MS and Lund, MN and Kiefer, R and Salabert, D and Garc{\'\i}a, RA and Davies, GR and Elsworth, Y and others},
  journal={The Astrophysical Journal Supplement Series},
  volume={237},
  number={1},
  pages={17},
  year={2018},
  publisher={IOP Publishing}
}

@ARTICLE{2010Sci...327..977B,
       author = {{Borucki}, William J. and {Koch}, David and {Basri}, Gibor and {Batalha}, Natalie and {Brown}, Timothy and {Caldwell}, Douglas and {Caldwell}, John and {Christensen-Dalsgaard}, J{\o}rgen and {Cochran}, William D. and {DeVore}, Edna and {Dunham}, Edward W. and {Dupree}, Andrea K. and {Gautier}, Thomas N. and {Geary}, John C. and {Gilliland}, Ronald and {Gould}, Alan and {Howell}, Steve B. and {Jenkins}, Jon M. and {Kondo}, Yoji and {Latham}, David W. and {Marcy}, Geoffrey W. and {Meibom}, S{\o}ren and {Kjeldsen}, Hans and {Lissauer}, Jack J. and {Monet}, David G. and {Morrison}, David and {Sasselov}, Dimitar and {Tarter}, Jill and {Boss}, Alan and {Brownlee}, Don and {Owen}, Toby and {Buzasi}, Derek and {Charbonneau}, David and {Doyle}, Laurance and {Fortney}, Jonathan and {Ford}, Eric B. and {Holman}, Matthew J. and {Seager}, Sara and {Steffen}, Jason H. and {Welsh}, William F. and {Rowe}, Jason and {Anderson}, Howard and {Buchhave}, Lars and {Ciardi}, David and {Walkowicz}, Lucianne and {Sherry}, William and {Horch}, Elliott and {Isaacson}, Howard and {Everett}, Mark E. and {Fischer}, Debra and {Torres}, Guillermo and {Johnson}, John Asher and {Endl}, Michael and {MacQueen}, Phillip and {Bryson}, Stephen T. and {Dotson}, Jessie and {Haas}, Michael and {Kolodziejczak}, Jeffrey and {Van Cleve}, Jeffrey and {Chandrasekaran}, Hema and {Twicken}, Joseph D. and {Quintana}, Elisa V. and {Clarke}, Bruce D. and {Allen}, Christopher and {Li}, Jie and {Wu}, Haley and {Tenenbaum}, Peter and {Verner}, Ekaterina and {Bruhweiler}, Frederick and {Barnes}, Jason and {Prsa}, Andrej},
        title = "{Kepler Planet-Detection Mission: Introduction and First Results}",
      journal = {Science},
         year = 2010,
        month = feb,
       volume = {327},
       number = {5968},
        pages = {977},
          doi = {10.1126/science.1185402},
       adsurl = {https://ui.adsabs.harvard.edu/abs/2010Sci...327..977B}
}

@ARTICLE{2017A&A...608A..87S,
       author = {{Salabert}, D. and {Garc{\'\i}a}, R.~A. and {Jim{\'e}nez}, A. and {Bertello}, L. and {Corsaro}, E. and {Pall{\'e}}, P.~L.},
        title = "{Photospheric activity of the Sun with VIRGO and GOLF. Comparison with standard activity proxies}",
      journal = {\aap},
         year = 2017,
        month = dec,
       volume = {608},
          eid = {A87},
        pages = {A87},
          doi = {10.1051/0004-6361/201731560},
archivePrefix = {arXiv},
       eprint = {1709.05110},
 primaryClass = {astro-ph.SR},
       adsurl = {https://ui.adsabs.harvard.edu/abs/2017A&A...608A..87S}
}

@ARTICLE{2016A&A...589A.118S,
       author = {{Salabert}, D. and {R{\'e}gulo}, C. and {Garc{\'\i}a}, R.~A. and {Beck}, P.~G. and {Ballot}, J. and {Creevey}, O.~L. and {P{\'e}rez Hern{\'a}ndez}, F. and {do Nascimento}, Jr., J. -D. and {Corsaro}, E. and {Egeland}, R. and {Mathur}, S. and {Metcalfe}, T.~S. and {Bigot}, L. and {Ceillier}, T. and {Pall{\'e}}, P.~L.},
        title = "{Magnetic variability in the young solar analog KIC 10644253. Observations from the Kepler satellite and the HERMES spectrograph}",
      journal = {\aap},
         year = 2016,
        month = may,
       volume = {589},
          eid = {A118},
        pages = {A118},
          doi = {10.1051/0004-6361/201527978},
archivePrefix = {arXiv},
       eprint = {1603.00655},
 primaryClass = {astro-ph.SR},
       adsurl = {https://ui.adsabs.harvard.edu/abs/2016A&A...589A.118S}
}

@ARTICLE{2018ApJ...852...46K,
       author = {{Karoff}, Christoffer and {Metcalfe}, Travis S. and {Santos}, {\^A}ngela R.~G. and {Montet}, Benjamin T. and {Isaacson}, Howard and {Witzke}, Veronika and {Shapiro}, Alexander I. and {Mathur}, Savita and {Davies}, Guy R. and {Lund}, Mikkel N. and {Garcia}, Rafael A. and {Brun}, Allan S. and {Salabert}, David and {Avelino}, Pedro P. and {van Saders}, Jennifer and {Egeland}, Ricky and {Cunha}, Margarida S. and {Campante}, Tiago L. and {Chaplin}, William J. and {Krivova}, Natalie and {Solanki}, Sami K. and {Stritzinger}, Maximilian and {Knudsen}, Mads F.},
        title = "{The Influence of Metallicity on Stellar Differential Rotation and Magnetic Activity}",
      journal = {\apj},
         year = 2018,
        month = jan,
       volume = {852},
       number = {1},
          eid = {46},
        pages = {46},
          doi = {10.3847/1538-4357/aaa026},
archivePrefix = {arXiv},
       eprint = {1711.07716},
 primaryClass = {astro-ph.SR},
       adsurl = {https://ui.adsabs.harvard.edu/abs/2018ApJ...852...46K}
}

@ARTICLE{1972ApJ...171..565S,
       author = {{Skumanich}, A.},
        title = "{Time Scales for Ca II Emission Decay, Rotational Braking, and Lithium Depletion}",
      journal = {\apj},
         year = 1972,
        month = feb,
       volume = {171},
        pages = {565},
          doi = {10.1086/151310},
       adsurl = {https://ui.adsabs.harvard.edu/abs/1972ApJ...171..565S}
}

@ARTICLE{2008ApJ...687.1264M,
       author = {{Mamajek}, Eric E. and {Hillenbrand}, Lynne A.},
        title = "{Improved Age Estimation for Solar-Type Dwarfs Using Activity-Rotation Diagnostics}",
      journal = {\apj},
         year = 2008,
        month = nov,
       volume = {687},
       number = {2},
        pages = {1264-1293},
          doi = {10.1086/591785},
archivePrefix = {arXiv},
       eprint = {0807.1686},
 primaryClass = {astro-ph},
       adsurl = {https://ui.adsabs.harvard.edu/abs/2008ApJ...687.1264M}
}

@ARTICLE{Appourchaux_2003,
       author = {{Appourchaux}, Thierry},
        title = "{Peak Bagging for Solar-like Stars}",
      journal = {\apss},
         year = 2003,
        month = mar,
       volume = {284},
       number = {1},
        pages = {109-119},
          doi = {10.1023/A:1023294327461},
       adsurl = {https://ui.adsabs.harvard.edu/abs/2003Ap&SS.284..109A}
}

@article{Nielsen_2021,
   title={PBjam: A Python Package for Automating Asteroseismology of Solar-like Oscillators*},
   volume={161},
   ISSN={1538-3881},
   url={http://dx.doi.org/10.3847/1538-3881/abcd39},
   DOI={10.3847/1538-3881/abcd39},
   number={2},
   journal={The Astronomical Journal},
   publisher={American Astronomical Society},
   author={Nielsen, M. B. and Davies, G. R. and Ball, W. H. and Lyttle, A. J. and Li 李坦, T. 达 and Hall, O. J. and Chaplin, W. J. and Gaulme, P. and Carboneau, L. and Ong 王加, J. M. J. 冕 and García, R. A. and Mosser, B. and Roxburgh, I. W. and Corsaro, E. and Benomar, O. and Moya, A. and Lund, M. N.},
   year={2021},
   month=jan, pages={62} }

@article{Nielsen_2025,
doi = {10.3847/1538-3881/adcb37},
url = {https://dx.doi.org/10.3847/1538-3881/adcb37},
year = {2025},
month = {may},
publisher = {The American Astronomical Society},
volume = {169},
number = {6},
pages = {322},
author = {Nielsen, M. B. and Ong, J. M. J. and Hatt, E. J. and Davies, G. R. and Chaplin, W. J. and Hookway, G. T. and Stokholm, A. and Scutt, O. J. and Lund, M. N. and García, R. A.},
title = {Asteroseismology with PBjam 2.0: Measuring Dipole Mode Frequencies in Coupling Regimes from Main-sequence to Low-luminosity Red Giant Stars},
journal = {The Astronomical Journal}
}

@INPROCEEDINGS{Harvey_1985,
       author = {{Harvey}, J.},
        title = "{High-Resolution Helioseismology}",
    booktitle = {Future Missions in Solar, Heliospheric \& Space Plasma Physics},
         year = 1985,
       editor = {{Rolfe}, Erica and {Battrick}, Bruce},
       series = {ESA Special Publication},
       volume = {235},
        month = jun,
        pages = {199},
       adsurl = {https://ui.adsabs.harvard.edu/abs/1985ESASP.235..199H}
}

@ARTICLE{Anderson_1990,
       author = {{Anderson}, Edwin R. and {Duvall}, Jr., Thomas L. and {Jefferies}, Stuart M.},
        title = "{Modeling of Solar Oscillation Power Spectra}",
      journal = {\apj},
         year = 1990,
        month = dec,
       volume = {364},
        pages = {699},
          doi = {10.1086/169452},
       adsurl = {https://ui.adsabs.harvard.edu/abs/1990ApJ...364..699A}
}

@article{Speagle_2020,
   title={dynesty: a dynamic nested sampling package for estimating Bayesian posteriors and evidences},
   volume={493},
   ISSN={1365-2966},
   url={http://dx.doi.org/10.1093/mnras/staa278},
   DOI={10.1093/mnras/staa278},
   number={3},
   journal={Monthly Notices of the Royal Astronomical Society},
   publisher={Oxford University Press (OUP)},
   author={Speagle, Joshua S},
   year={2020},
   month=feb, pages={3132–3158}
}

@article{Foreman_Mackey_2013,
   title={<tt>emcee</tt>: The MCMC Hammer},
   volume={125},
   ISSN={1538-3873},
   url={http://dx.doi.org/10.1086/670067},
   DOI={10.1086/670067},
   number={925},
   journal={Publications of the Astronomical Society of the Pacific},
   publisher={IOP Publishing},
   author={Foreman-Mackey, Daniel and Hogg, David W. and Lang, Dustin and Goodman, Jonathan},
   year={2013},
   month=mar, pages={306–312}
}

@misc{Hookway_2025,
      title={Peakbagging the K2 KEYSTONE sample with PBjam: characterising the individual mode frequencies in solar-like oscillators}, 
      author={George T. Hookway and Martin B. Nielsen and Guy R. Davies and Mikkel N. Lund and Rafael A. García and Savita Mathur and Victor See and Amalie Stokholm},
      year={2025},
      eprint={2510.21626},
      archivePrefix={arXiv},
      primaryClass={astro-ph.SR},
      url={https://arxiv.org/abs/2510.21626}, 
}

@ARTICLE{2025A&A...700A..39K,
       author = {{Kjeldsen}, Hans and {Bedding}, Timothy R. and {Li}, Yaguang and {Grundahl}, Frank and {Andersen}, Mads Fredslund and {Wright}, Duncan J. and {Soutter}, Jack and {Wittenmyer}, Robert and {Reyes}, Claudia and {Stello}, Dennis and {Crawford}, Courtney and {Zhou}, Yixiao and {Clerte}, Mathieu and {Pall{\'e}}, Pere L. and {Simon-Diaz}, Sergio and {Christensen-Dalsgaard}, J{\o}rgen and {Handberg}, Rasmus and {Hansen}, Hasse and {Heeren}, Paul and {Jessen-Hansen}, Jens and {Lund}, Mikkel N. and {Lundkvist}, Mia S. and {Brogaard}, Karsten and {Tronsgaard}, Ren{\'e} and {Rudrasingam}, Jonatan and {Casagrande}, Luca and {Horner}, Jonathan and {Huber}, Daniel and {Lattanzio}, John and {Martell}, Sarah L. and {Murphy}, Simon J.},
        title = "{Asteroseismology of the G8 subgiant {\ensuremath{\beta}} Aquilae with SONG-Tenerife, SONG-Australia and TESS}",
      journal = {\aap},
         year = 2025,
        month = aug,
       volume = {700},
          eid = {A39},
        pages = {A39},
          doi = {10.1051/0004-6361/202554633},
archivePrefix = {arXiv},
       eprint = {2506.00493},
 primaryClass = {astro-ph.SR},
       adsurl = {https://ui.adsabs.harvard.edu/abs/2025A&A...700A..39K}
}

% Alternatively you could enter them by hand, like this:
% This method is tedious and prone to error if you have lots of references
%\begin{thebibliography}{99}
%\bibitem[\protect\citeauthoryear{Author}{2012}]{Author2012}
%Author A.~N., 2013, Journal of Improbable Astronomy, 1, 1
%\bibitem[\protect\citeauthoryear{Others}{2013}]{Others2013}
%Others S., 2012, Journal of Interesting Stuff, 17, 198
%\end{thebibliography}

%%%%%%%%%%%%%%%%%%%%%%%%%%%%%%%%%%%%%%%%%%%%%%%%%%

%%%%%%%%%%%%%%%%% APPENDICES %%%%%%%%%%%%%%%%%%%%%

\appendix

%%%%%%%%%%%%%%%%%%%%%%%%%%%%%%%%%%%%%%%%%%%%%%%%%%
% Don't change these lines
\bsp	% typesetting comment
\label{lastpage}
\end{document}